\documentclass[aps,twocolumn,amsmath,amssymb,superscriptaddress,longbibliography,showpacs,nofootinbib,numbers,openany]{revtex4-1}

\usepackage{color}
\usepackage{latexsym}
\usepackage{float}
\usepackage{dsfont}
\usepackage{accents} 
\usepackage{bibunits}
\usepackage{url}

\usepackage{bm, mathtools}
\usepackage{graphicx}
\usepackage{dcolumn}
\usepackage{times}
\usepackage[english]{babel}
\usepackage[T1]{fontenc}
\usepackage{xcolor}

\usepackage{cellspace}%
\usepackage{makecell}
\setcellgapes{3pt}

\newcommand{\tcr}[1]{\textcolor{black}{#1}}

\definecolor{linkcolor}{rgb}{0.2,0.2,1} 
\usepackage[pdftex,colorlinks=true, pdfstartview=FitV, linkcolor= linkcolor, citecolor= linkcolor, urlcolor= linkcolor, hyperindex=true,hyperfigures=true]{hyperref} 

\newcommand{\mean}[1]{\overline{#1}}

\newcommand{\supplementarysection}{%
  \setcounter{figure}{0}
  \let\oldthefigure\thefigure
  \renewcommand{\thefigure}{S\oldthefigure}
  \section{Supplementary section}
  \let\oldchapter\chapter
  \renewcommand{\chapter}{
    \let\thefigure\oldthefigure
    \let\chapter\oldchapter
    \oldchapter
  }
}

\begin{document}

\title{Variance Sum Rule for Entropy Production}

\author{I. Di Terlizzi$^{*}$}

\address{Max Planck Institute for the Physics of complex systems, N{\"o}thnitzer Stra{\ss}e 38, 01187, Dresden, Germany} 
\address{Dipartimento di Fisica e Astronomia, Universit\`a di Padova, Via Marzolo 8, 35131, Padova, Italy
} 

\author{M. Gironella$^{*}$}

\address{Small Biosystems Lab, Condensed Matter Physics Department, Universitat de Barcelona, C/ Marti i Franques 1, 08028 Barcelona, Spain}

\address{Department of Medical Biochemistry and Cell Biology, Institute of Biomedicine, The Sahlgrenska Academy, University of Gothenburg, 40530 Gothenburg, Sweden}

\author{D. Herraez-Aguilar}

\address{Facultad de Ciencias Experimentales, Universidad Francisco de Vitoria, Ctra. Pozuelo-Majadahonda Km 1,800, 28223, Pozuelo de Alarcón, Madrid, Spain}

\author{T. Betz}

\address{Third Institute of Physics, Georg August Universit{\"a}t G{\"o}ttingen, G{\"o}ttingen, Germany}

\author{F. Monroy}

\address{Departamento de Qu\'{\i}mica F\'{\i}sica, Facultad de Qu\'{\i}mica, Universidad Complutense, 28040 Madrid, Spain}

\address{Translational Biophysics, Instituto de Investigaci\'on Sanitaria Hospital Doce de Octubre (IMAS12), Av. Andaluc\'{\i}a, 28041 Madrid, Spain}

\author{M. Baiesi}

\address{Dipartimento di Fisica e Astronomia,
 Universit\`a di Padova, Via Marzolo 8, 35131, Padova, Italy
} 
\address{INFN, Sezione di Padova, Via Marzolo 8, 35131, Padova,
Italy}

\author{F. Ritort}

\address{Small Biosystems Lab, Condensed Matter Physics Department, Universitat de Barcelona, C/ Marti i Franques 1, 08028 Barcelona, Spain}

\address{Institut de Nanoci\`encia i Nanotecnologia (IN2UB), Universitat de Barcelona, 08028 Barcelona, Spain}

\date{\today}

\begin{abstract} 
Entropy production is the hallmark of nonequilibrium physics, quantifying irreversibility, dissipation, and the efficiency of energy transduction processes. Despite many efforts, its measurement at the nanoscale remains challenging. We introduce a variance sum rule for displacement and force variances that permits us to measure the entropy production rate $\sigma$ in nonequilibrium steady states. We first illustrate it for directly measurable forces, such as an active Brownian particle in an optical trap. We then apply the variance sum rule to flickering experiments in human red blood cells. We find that $\sigma$ is spatially heterogeneous with a finite correlation length and its average value agrees with calorimetry measurements. The VSR paves the way to derive $\sigma$ using force spectroscopy and time-resolved imaging in living and active matter.
 \end{abstract}

\maketitle

 Nonequilibrium steady states (NESS) pervade nature, from climate dynamics \cite{singh2022climate} to living cells and active matter \cite{bechinger2016active}. A fundamental quantity is the entropy production rate $\sigma$ at which energy is dissipated to the environment, which is positive by the second law of thermodynamics \cite{maes03_1,seifert2012stochastic}. Entropy production measurements remain challenging despite their relevance, especially in microscopic systems with stochastic and spatially varying fluctuations and limited access to microscopic variables \cite{ritort2008nonequilibrium,ciliberto2017experiments}. The entropy production rate $\sigma$ determines the efficiency of energy transduction in classical and quantum systems \cite{martinez2016brownian,landi2021irreversible}, the energetic costs and \tcr{irreversible behavior of} living cells \cite{martin2001comparison,battle2016broken,turlier2016equilibrium,lynn2021broken}. It is an elusive quantity when forces and currents are experimentally inaccessible.  Bounds can be obtained from time irreversibility \cite{li2019quantifying,roldan2021quantifying}, the thermodynamic uncertainty relation~\cite{bar15,horowitz2020thermodynamic} and coarse-graining  ~\cite{bisker2017hierarchical,teza2020exact,skinner2021improved,dechant2021improving,dieball2022mathematical}. \tcr{Most of these results provide lower bounds that refine the second law of thermodynamics, $\sigma\ge 0$. However, the bounds are often loose without upper limits and therefore uninformative about the actual $\sigma$. Alternative methods that estimate $\sigma$ more precisely are needed to determine dissipative processes in the nanoscale. } 

\begin{figure}[!t] 
   \centering
   \includegraphics[width=0.48\textwidth]{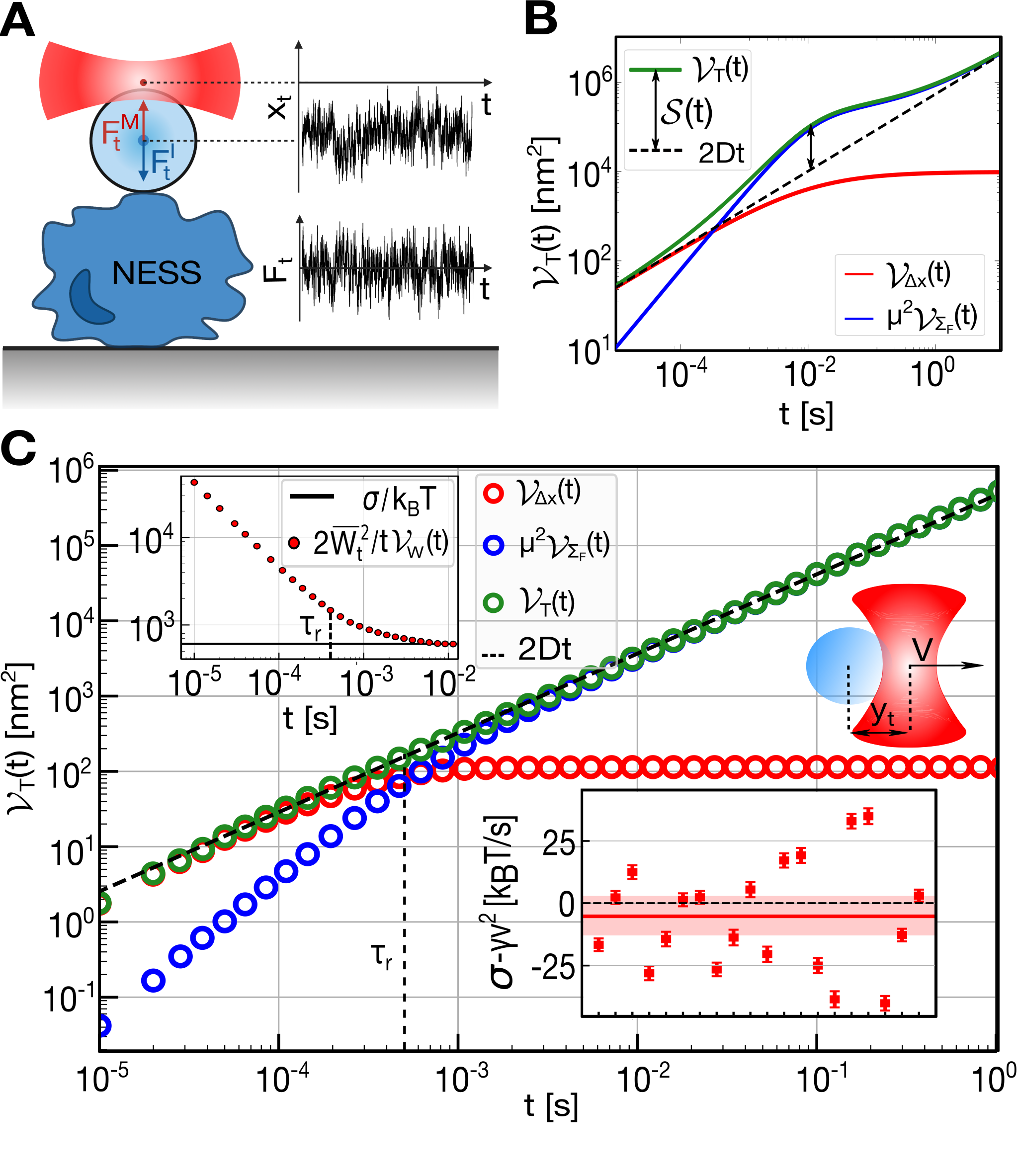} 
   \caption{
   {\bf Variance sum rule (VSR): sketches and experiments with dragged particle.}
   ({\bf A}) Experimental setup for a NESS measured with optical tweezers.
   ({\bf B}) Illustration of the VSR showing the different terms in \eqref{eq:VSR1}.
   ({\bf C}) Experimental test of the VSR for an optically trapped bead dragged through water at room temperature (bead radius $R=1.5\mu m$, mobility $\mu=4     \cdot 10^4$nm/pN$\cdot$s, speed $v=10\mu m/s$, $\gamma v^2=610 k_BT/s$). The lower inset plots  $\sigma-\gamma v^2=\frac{1}{4\mu}\partial_{t}^2{\cal V}_{\Delta x}|_{t=0}+\frac{\mu}{2} {\cal V}_{F}$ from \eqref{eq:EP1} for the experimental realizations; the horizontal red line shows the average over all experiments ($-5(7) k_B T/s$) with one standard deviation (red band). The black dashed line is the theoretical prediction $\sigma-\gamma v^2=0$. The upper inset shows the experimental test of the inequality \eqref{eq:RTUR}. Dashed vertical lines show the bead's relaxation time $\tau_r$.}
   \label{fig:FIG1}
\end{figure}
\noindent
{\bf\large \tcr{Variance sum rule}}\\
\noindent
We introduce a variance sum rule (VSR) to derive $\sigma$ in experiments where a measurement probe is in contact with a system in a NESS (Fig.~\ref{fig:FIG1}A). Dynamics is described by a Langevin equation, $\dot{x}(t)=\mu F_t+\sqrt{2D}\eta_t$, with probe mobility $\mu$, diffusivity $D$ and $\eta_t$ a Gaussian white noise. The total force acting on the probe $F_t\equiv F_t(x_t)$ equals the sum of the force exerted by the measurement device, $F_t^M$, plus a probe-system interaction, $F_t^I$, $F_t=F_t^M+F_t^I$ (arrows in Fig.~\ref{fig:FIG1}A). In most experimental settings, $F_t^I$ \tcr{remains inaccessible}, so $F_t$ and $\sigma$ cannot be directly measured. Our approach focuses on how observables ${Q_t}$ on average spread in time, as quantified by their variance ${\cal V}_{Q}(t)=\overline{Q^2_t}-\overline{Q_t}^2$ with \tcr{$\overline{(\dots)}$} the dynamical average in the NESS. The VSR is an equality for integrated quantities in an arbitrary time interval (0,$t$), which imposes a tight constraint on the fluctuations in a stochastic diffusive system over the experimental timescales. By integrating the Langevin equation on the interval (0,$t$) and by taking the variance of both sides, a time-preserved identity can be obtained (Sec. \ref{S1:VSRDer}, Supp. Mat.). The VSR for position and force fluctuations reads
\begin{equation}
   {\cal V}_{\Delta x}(t)+\mu^2{\cal V}_{\Sigma_ F}(t)=2Dt+{\cal S}(t) \,,
\label{eq:VSR1}
\end{equation}
where the lhs includes the variances of the displacements $\Delta x_t=x_t-x_0$, and of time-cumulative forces ($\Sigma_F(t)=\int_0^t ds F_s$). The  total variance ${\cal V}_T(t)={\cal V}_{\Delta x}(t)+\mu^2{\cal V}_{\Sigma_ F}(t)$ equals the free  diffusion term $2Dt$ plus a nonequilibrium contribution ${\cal S}(t)$ denoted as excess variance, 
\begin{equation}
   {\cal S}(t)=2\mu\int_0^t d s
   \left[C_{x F}(s) - C_{F x}(s) \right]
\label{eq:S1}
\end{equation}
that measures the breakdown of time-reversal symmetry, with $C_{A B}(s)=\overline{A_s B_0}-\overline{A}_s\,\overline{B}_0$ the correlation function in the NESS. 
In equilibrium, ${\cal S}(t)=0$ due to time-reversal symmetry.
Figure \ref{fig:FIG1}B illustrates the VSR for a generic NESS.

From the VSR, one can derive an equation relating $\sigma$ to the variances of fluctuating variables. By taking the time derivative twice of \eqref{eq:S1} and evaluating it at $t=0$, one obtains a formula for $\sigma$ that depends on the convexity of the excess variance ${\cal S}(t)$ at $t=0$ (Sec.\ref{SS2:EPfromVSR}, Supp. Mat.),
\begin{equation}
  \sigma=\frac{v^2}{\mu}+\frac{1}{4\mu}\partial_{t}^2 {\cal S}|_{t=0}
\label{eq:EP0}
\end{equation}
where $v=\overline{\dot x}$ is the particle's average velocity and $\sigma$ is expressed in power units (e.g. $k_BT/s$). By using \eqref{eq:VSR1} along with \eqref{eq:EP0}, we derive the formula for the rate of entropy production in terms of the static variance of the force ${\cal V}_{F}=\overline{F^2}-\overline{F}^2$ and the convexity of the mean-squared displacement ${\cal V}_{\Delta x}$ at time 0, 
\begin{equation}
      \sigma
      =\frac{v^2}{\mu}+\frac{1}{4\mu}\partial_{t}^2{\cal V}_{\Delta x}|_{t=0}+\frac{\mu}{2} {\cal V}_{F}\,.
\label{eq:EP1}
\end{equation}
To illustrate the VSR, we consider two examples of a NESS where $F_t$ equals the force in the measurement device, $F_t=F_t^M$, and $F_t^I=0$.  \\

\noindent
{\bf\large \tcr{Bead dragged through water}}\\
\noindent
The first \tcr{system} is an optically trapped colloidal particle dragged through water (friction coefficient $\gamma=1/\mu$) at speed $v$. Bead's dynamics can be analytically solved, and the VSR \eqref{eq:VSR1} verified (Sec.S3, Supp. Mat.).
Equation \eqref{eq:EP1} follows with ${\cal S}=0$ and $\sigma=\gamma v^2$, as expected. Figure \ref{fig:FIG1}C shows the experimental validation of the VSR \eqref{eq:VSR1}. The right inset shows measurements of $\sigma-\gamma v^2$ for several \tcr{repetitions of the experiment and} using \eqref{eq:EP1}, \tcr{finding} $\sigma-\gamma v^2=-5(7) k_B T/s$. Notice that ${\cal S}=0$ implies that the two rightmost terms in \eqref{eq:EP1} are of equal magnitude but opposite sign compensating each other, $\mu {\cal V}_{F}=-\frac{1}{2\mu}\partial_{t}^2{\cal V}_{\Delta x}|_{t=0}=k_BT/\tau_r>0$ with $\tau_r=\gamma/k=0.35ms$ the bead's relaxation time ($k=70pN/\mu m$ being the trap stiffness). The value $\mu {\cal V}_{F}\sim 3\cdot 10^3 k_BT/s$ is almost three orders of magnitude larger than $\sigma-\gamma v^2$ ($\pm 7 k_BT/s$).

The results ${\cal S}=0$ and $\sigma=\gamma v^2$ are not restricted to a harmonic well but hold for an arbitrary time-dependent potential $U(x-vt)$. 
This gives a reversed thermodynamic uncertainty relation \cite{horowitz2020thermodynamic} for the work exerted on the bead by the optical trap, $W_t=v\Sigma_F(t)=v\int_0^tdsF_s$, and an upper bound for $\sigma$ (Sec.S4, Supp. Mat.),
\begin{equation}
   \frac{\sigma}{k_BT}\le  \frac{2\overline{W_t}^2}{t\,{\cal V}_{W}(t)}\,.
   \label{eq:RTUR}
\end{equation}
In Fig~\ref{fig:FIG1}C (left inset), we experimentally test \eqref{eq:RTUR}. The upper bound becomes tight for $t\gg\tau_r$, the difference between two terms in \eqref{eq:RTUR} vanishing like $\tau_r/t$, as expected from the steady-state fluctuation theorem for Gaussian work distributions \cite{seifert2012stochastic}.\\
\begin{figure*}[t!] 
   \centering
   \includegraphics[width=\linewidth]{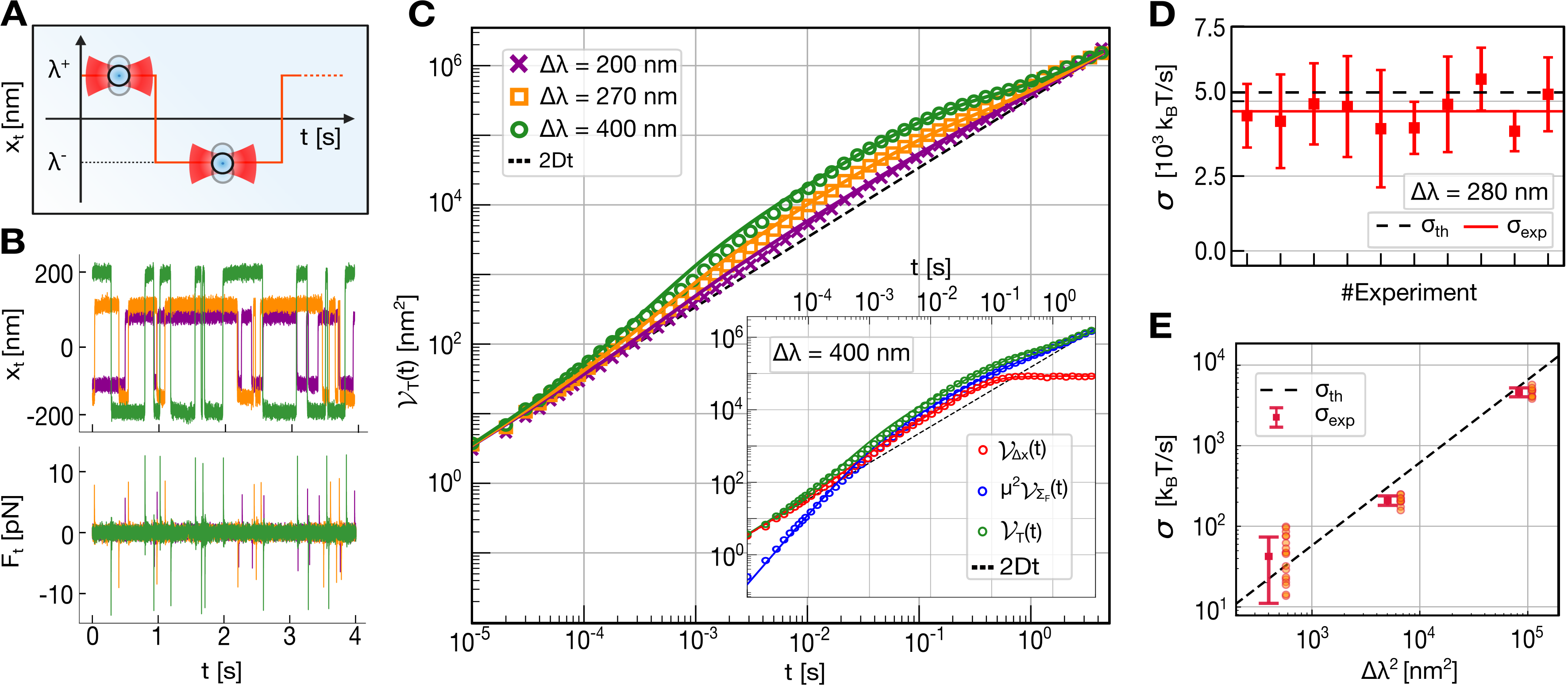} 
   \caption{{\bf VSR and entropy production rate for experiments with a stochastic switching trap}. ({\bf A}) Schematics of the experiment. ({\bf B}) Traces of position and force for three $\Delta\lambda$ values (see legend in  C).  ({\bf C}) VSR \eqref{eq:VSR1} and total variance ${\cal V}_T$: symbols are experimental data, and lines represent the theory with known parameters without fitting. The inset shows the different terms in the VSR. ({\bf D})  Measurements of $\sigma$ for $w_+=w_-=10 s^{-1}$ and $\Delta\lambda=280nm$; we show different experimental realizations (squares), their average $\sigma_{\text{exp}}$ and the theoretical value $\sigma_{\text{th}}$ \eqref{eq:SWT2}. 
   ({\bf E}) $\sigma$ (red symbols) averaged over experimental realizations (orange circles) for  $\Delta\lambda=18nm,70nm,280nm$; black line is the analytical prediction \eqref{eq:SWT2}.\\\\ }
   \label{fig:FIG2}
\end{figure*}

\noindent
{\bf\large \tcr{The stochastic switching trap}}\\
\noindent
The second \tcr{system} we consider is the stochastic switching trap (SST) \cite{dieterich2015single}, where an active force is applied to an optically trapped bead by randomly switching the trap position $\lambda_t$ between two values ($\lambda_+,\lambda_-$) separated by $\Delta \lambda=\lambda_+-\lambda_-$ (Fig.~\ref{fig:FIG2}A). Jumps occur at exponentially distributed times with switching rates $w_+,w_-$ at each position. The ratio $w_-/w_+=q/(1-q)$ defines the probability $q$ of the trap to be at position $\lambda_+$. Figure \ref{fig:FIG2}B shows the measured bead's position $x_t$ and force $F_t=k(\lambda_t-x_t)$ for three cases with $q=1/2$ and varying $\Delta \lambda$. The bead follows the movement of the trap (top), quickly relaxing to its new equilibrium trap position at every jump (force spikes, bottom).  Figure \ref{fig:FIG2}C shows the total variance, ${\cal V}_T(t)={\cal V}_{\Delta x}(t)+\mu^2{\cal V}_{\Sigma_ F}(t)$. ${\cal V}_T$ deviates from $2Dt$ (dashed line) between $10^{-4}$s and 1s, showing that ${\cal S}\ne 0$ is comparable to ${\cal V}_T$ (notice the log-log scale). The SST model is analytically solvable (Sec.S5, Supp. Mat.), giving expressions for ${\cal V}_{\Delta x}(t),{\cal V}_{\Sigma_ F}(t)$ and ${\cal S}(t)$. For the latter, we find
\begin{equation}
{\cal S}(t)=4(\Delta\lambda)^2q(1-q)\frac{ \alpha(1-e^{-w_r t})-\alpha^2(1-e^{-w t})}{1-\alpha^2}
\label{eq:SWT1}
\end{equation}
with $w=w_++w_-$, $\alpha=w_r/w$, and $w_r=1/\tau_r=k/\gamma$ (bead's relaxation rate for a resting trap). In Fig.~\ref{fig:FIG2}C we test the VSR and \eqref{eq:SWT1} for three NESS conditions. The inset shows the two terms contributing to the total variance ${\cal V}_T$. 
For large times, ${\cal S}$ converges to a finite value, and ${\cal V}_T$ merges with the equilibrium line $2Dt$ (black dashed line) when plotted in log-log scale. 
Eqs.\eqref{eq:EP0},\eqref{eq:SWT1} yield the theoretical prediction ($v=0$)
\begin{equation}
\sigma_{\rm th}= (k\Delta\lambda)^2q(1-q)\mu \frac{ w}{w+w_r}\,.
\label{eq:SWT2}
\end{equation}
Figure \ref{fig:FIG2}D shows values of $\sigma$ measured in SST experiments with $\Delta\lambda=280nm$ using \eqref{eq:EP1}. Their average $\sigma_{\rm exp}=4.6(4)\cdot 10^3 k_BT/s$ agrees with the theoretical prediction \eqref{eq:SWT2}, $\sigma_{\rm th}\sim 5.3\cdot 10^3  k_BT/s$. 
Figure \ref{fig:FIG2}E compares $\sigma_{\rm exp}$ with $\sigma_{\rm th}$ \eqref{eq:SWT2} (black dashed line) for varying $\Delta\lambda$. Experiment and theory agree over three decades of $\sigma$.\\

\noindent
{\bf\large\tcr{Reduced-VSR}}\\
\noindent
Until now, we have considered the case of a single degree of freedom where the total force acting on the bead equals the measured force, $F_t=F_t^M$ and $F_t^I=0$. For the case of multiple degrees of freedom where positions and total forces can be measured, Eqs.\eqref{eq:VSR1} and \eqref{eq:EP0}, can be generalized (Sec. S1 and S2, Supp. Mat.). 
Quite often, however, a measurement probe (AFM tip, microbead, etc.) is in contact with a system in a NESS, such as a biological cell with metabolic activity, Fig.~\ref{fig:FIG1}A. In this case, $F_t^I\ne 0$ is experimentally inaccessible, and $F_t=F_t^M+F_t^I$ cannot be measured, making the VSR \eqref{eq:VSR1} inapplicable. Moreover, in many cases, only a spatial degree of freedom $x_t$ is monitored, e.g., in particle tracking experiments~\cite{manzo2015review,scott2023extracting}, or in detecting cellular fluctuations \cite{ahmed2018active,salinas2022membrane}.  To apply the VSR in these situations, it is necessary to model the NESS by making assumptions about the interaction $F_t^I$ and the underlying degrees of freedom. Specifically, for a linear-response measuring device ($F_t^M=-kx_t$), a reduced-VSR for a single degree of freedom can be derived and expressed in terms of variances related to the position $x_t$ only. In these conditions, the displacement variance, ${\cal V}_{\Delta x}$, along with the variance of $\Sigma_x (t)=\int^t_0ds\, x_s$, ${\cal V}_{\Sigma_x}(t)$, satisfy (Sec.S6, Supp. Mat.),
\begin{equation}
   {\cal V}_{\Delta x}(t)+\mu^2k^2\,{\cal V}_{\Sigma_x}(t)=2Dt+{\cal \tilde{S}}(t) \,.
\label{eq:VSRABP}
\end{equation}
Equation \eqref{eq:VSRABP} is a general result which, however, does not permit to derive a formula for $\sigma$ like \eqref{eq:EP0}. Notice that ${\cal \tilde{S}}$ differs from ${\cal{S}}$ in \eqref{eq:VSR1} and does not vanish in equilibrium.  ${\cal \tilde{S}}$ can be expressed in terms of the generic interacting force $F_t^I$, see Eq.\eqref{eq:redS_comp} in Supp. Mat. To derive $\sigma$ using \eqref{eq:VSRABP}, we use a solvable model for the experiment and a procedure consisting of the following steps: 1) Analytically derive expressions for the excess variances, ${\cal S}(t)$ and ${\cal \tilde{S}}(t)$ for the model; 2) Calculate $\sigma_{\rm th}$ from ${\cal S}(t)$ using (3); 3) Fit the reduced-VSR \eqref{eq:VSRABP} to the experimental data using ${\cal \tilde{S}}(t)$ from the model to extract the model parameters; 4) Insert the model parameters in the analytical expression for $\sigma_{\rm th}$ to derive $\sigma$. We stress that for solvable models with multiple degrees of freedom, \eqref{eq:VSR1} and \eqref{eq:EP0} can be generalized (Eq. \eqref{eq:VSR_multi} and \eqref{eq:sigma-d}, Supp. Mat.). The approach remains applicable to a vast category of NESS whenever the interacting force $F_I$ between the probe and NESS is linear. This is a typical situation in mesoscopic systems where fluctuations are small in the linear response regime. A model for the experimental system that includes the degrees of freedom contributing most to $\sigma$ is required. 
\begin{figure*}
   \centering
   \includegraphics[width=\linewidth]{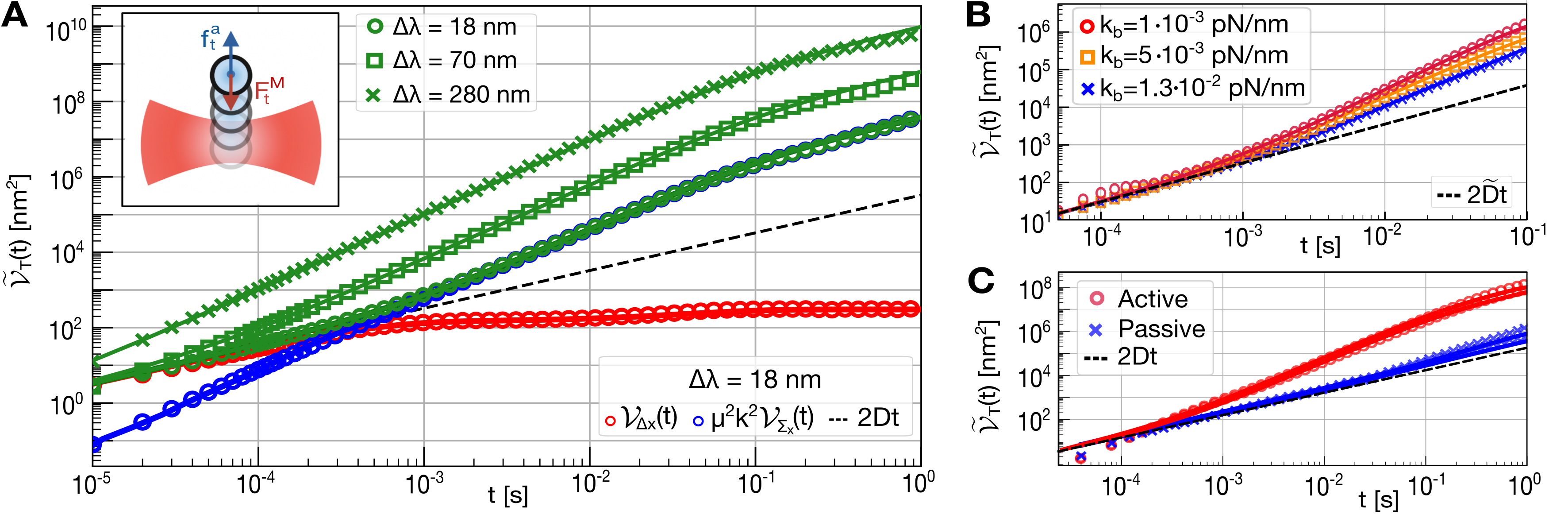} 
   \caption{{\bf Application of the reduced-VSR to experiments (SST and RBCs) to extract the entropy production rate.} ({\bf A}) Test of \eqref{eq:VSRABP} for the SST experimental data, equivalent to the active Brownian particle (ABP), in a harmonic trap (Eq.\eqref{eq:ABP} and inset).
    Symbols are experimental values for  $\tilde{\cal V}_{T}(t) = {\cal V}_{\Delta x}(t)+\mu^2k^2\,{\cal V}_{\Sigma_x}(t)$, fitted to \eqref{eq:VSRABP} for different $\Delta\lambda$ (lines).
    Blue and red circles are the two contributions to $\tilde{\cal V}_{T}(t)$ for $\Delta\lambda=18nm$. ({\bf B, C})  Fits of the reduced-VSR to $\tilde{\cal V}_{T}(t)$ for the two-layer active model (Sec.\ref{S6:RedVSR}, Supp. Mat.). Panel B: healthy RBCs in OT-stretching experiments at three forces (Fig.~\ref{fig:FIG4}A). \tcr{To help visualization, the three different $\tilde{\cal V}_{T}(t)$ have been scaled with respect to a single $\tilde{D}$ value}; Panel C: healthy and passive RBCs in OT-sensing experiments (Fig.~\ref{fig:FIG4}B).
   }
   \label{fig:FIG3}
\end{figure*}
For instance, let us consider an active Brownian particle (ABP) in an optical trap subject to a random time-correlated active force $F_t^I\equiv f_t^a$ of amplitude $\epsilon$, $\overline{f_t^a}=0$, $\overline{f_t^af_s^a}=\epsilon^2 e^{-|t-s|/\tau_a}$, with $\tau_a$ the active correlation time (Fig.~\ref{fig:FIG3}A, inset). Dynamics are described by the stochastic equation,
\begin{equation}
    \dot x_t =-k\mu x_t + \sqrt{2D}\eta_t + \mu f_t^a
    \label{eq:ABP}
\end{equation}
with $k$ the trap stiffness, $\mu$ the particle mobility, and $D=k_BT\mu$ the diffusion constant. To test the reduced-VSR approach \eqref{eq:VSRABP} for deriving $\sigma$, we exploit the mapping of the ABP \eqref{eq:ABP} to the SST model discussed previously (Fig.~\ref{fig:FIG2}A). The mapping follows by identifying parameters $\epsilon=k\Delta\lambda \sqrt{q(1-q)}$, $\tau_a=1/w$, $w_r=k\mu$ from which \eqref{eq:SWT2} follows (for $q=1/2$, see also~\cite{garcia2021run}). 
We have used \eqref{eq:VSRABP} to analyze the data already used in the previous approach for the SST experiments (Fig.~\ref{fig:FIG2}) with 
${\cal \tilde{S}}(t)=2\epsilon^2\mu^2\tau_a[t-\tau_a(1-e^{-t/\tau_a})]$ (c.f. Eq.\eqref{eq:VSR_ABP_supp}) where $\epsilon$ and $\tau_a$ are fitting parameters. Results are shown in Fig.~\ref{fig:FIG3}A and residuals in Fig.~\ref{fig:S2}A. Their values and $\sigma$ agree with the expected ones (Table I and Fig.~\ref{fig:S3}A in Supp. Mat.). Therefore, the reduced-VSR \eqref{eq:VSRABP} permits us to infer NESS parameters and $\sigma$ from $x_t$ measurements only.\\
\begin{figure*}[t!] 
   \centering
   \includegraphics[width=\linewidth]{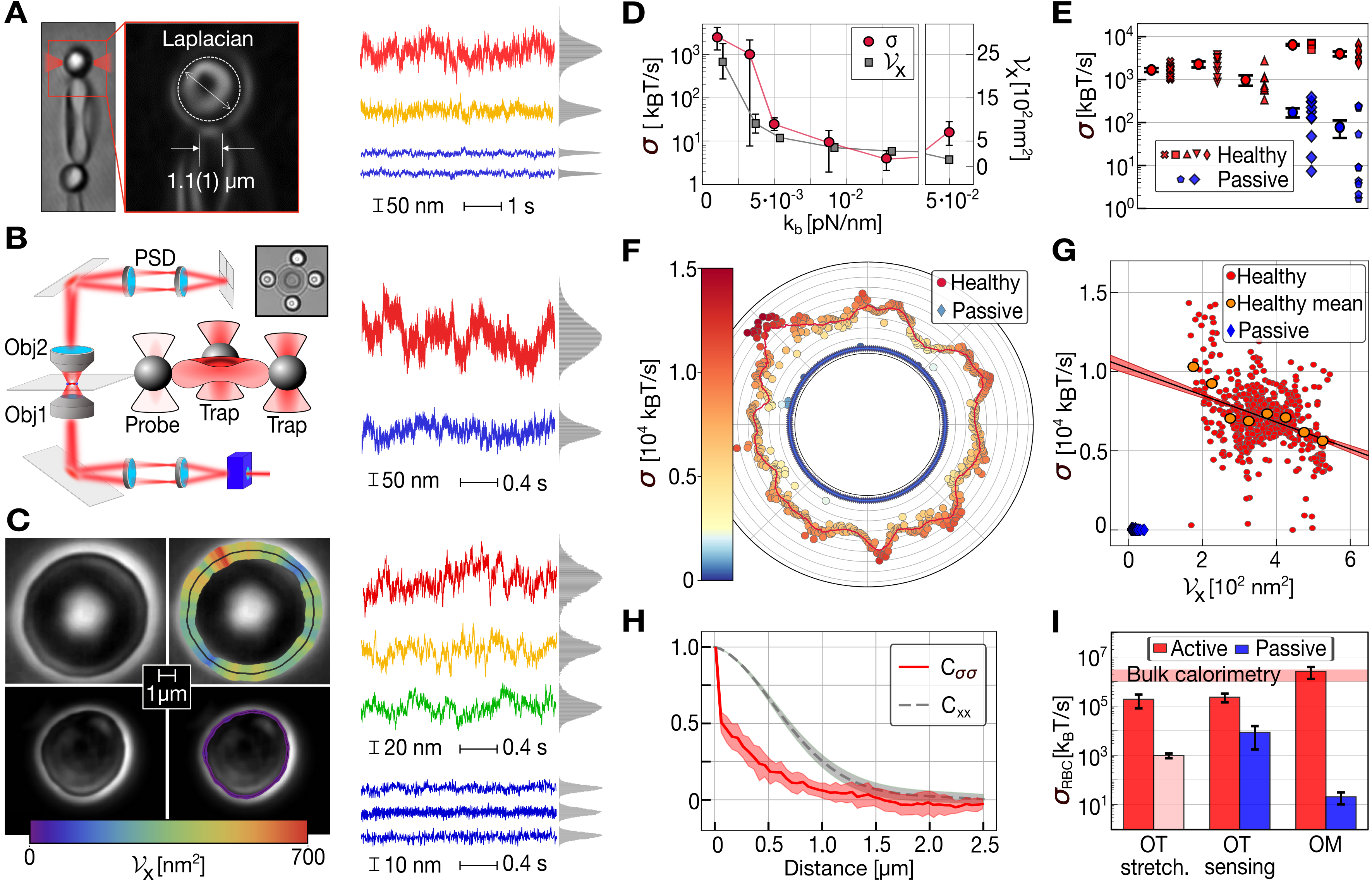} 
   \caption{{\bf Application of the reduced-VSR to RBCs.} ({\bf A}) OT-stretching experiments. Video image of stretched RBC and schematics of contact area estimation (left); \tcr{(right) three selected bead position traces at a high (blue), medium (orange), and low (red) trap stiffness.} ({\bf B}) OT-sensing experiments. Experimental setup from Ref.\cite{turlier2016equilibrium} (left) and tracking bead position traces for a healthy (red) and passive (blue) RBC. ({\bf C}) Ultrafast OM measurements: healthy RBC (upper images) and position traces (right) for three selected pixels (50nm$\times$50nm) along the cell contour with high (red), medium (yellow), and low (green) variance ${\cal V}_x$; passive RBC (lower images) and cell contour traces for three selected pixels (blue, right). The right images also show a color variance map along the cell contour. The color bar denotes variance levels (red, highest; blue, lowest). ({\bf D}) \tcr{$\sigma$ and position variance ${\cal V}_x$ measurements for OT-stretching (A) varying the trap stiffness $k_b$ from high values ($5\cdot 10^{-2}$pN/nm, rightmost points) to low values ($7\cdot 10^{-4}$pN/nm, leftmost points) for healthy RBCs}. ({\bf E}) $\sigma$ measurements for OT-sensing for healthy (red symbols) and passive (blue symbols) RBCs.({\bf F}) Colored $\sigma$-map for OM measurements  along the equatorial cell contour, as in panel C, for a healthy RBC (circles) and a passive RBC (diamonds). The radial distance represents $\sigma$ in a.u. The orange curve is the $\sigma$-smoothed profile. ({\bf G}) Scatter plot of $\sigma$ versus ${\cal V}_x$ for the RBCs of panel F, showing they are partially anticorrelated. Orange circles are $\sigma$ values averaged over windows of 50 nm$^2$ in ${\cal V}_x$. ({\bf H}) Spatial correlation functions for $\sigma$ and position $x$ are measured along the cell contour. ({\bf I}) Values of $\sigma_{\rm RBC}$ compared to calorimetry estimates. \tcr{For OT-stretching, the dark (light) red bar corresponds to the lowest (highest) trap stiffness.}}
   \label{fig:FIG4}
\end{figure*}

\noindent
{\bf\large \tcr{Red blood cells}}\\
\noindent
Finally, we apply the reduced-VSR to the challenging case of human red blood cells (RBCs) \cite{turlier2019unveiling}. RBCs metabolize glucose into ATP via the glycolytic pathway, producing the cell membrane's active flickering with a consequent entropy creation \cite{yoon2009flickering,betz2009atp,rodriguez2015direct,turlier2016equilibrium}. The RBC membrane is dynamically attached to the spectrin cortex via multiprotein complexes, which actively bind and unbind in the phosphorylation step of the glycolytic pathway \cite{byers1985visualization}. We have carried out experimental RBC measurements using three techniques (Fig.~\ref{fig:FIG4}). Two of them use optical tweezers (OT) in different setups: {\it i) mechanical stretching} of RBCs using beads non-specifically attached to the membrane \tcr{with different optical trap stiffness} (OT-stretching, Fig.~\ref{fig:FIG4}A); {\it ii) mechanical sensing} of a biotinylated RBC membrane using streptavidin functionalized beads using data from \cite{turlier2016equilibrium} (OT-sensing, Fig.~\ref{fig:FIG4}B). The third technique measures cell contour fluctuations by membrane flickering  segmentation  tracking of free-standing RBCs using ultrafast optical microscopy (OM)~\cite{rodriguez2015direct,mell2018gradient} (Fig.~\ref{fig:FIG4}C). 
As a first observation, a single-layer active model \eqref{eq:ABP} with its ${\cal \tilde{S}}(t)$ in \eqref{eq:VSRABP} does not describe the experimental data. Instead,
we consider a two-layer model with one hidden position variable for the active membrane-cortex interaction that is linearly coupled to the membrane outer layer $x$ (probe) (Sec.\ref{S7:RBCmodel}, Supp. Mat.). Similar active models have been proposed in the study of hair-cell bundle dynamics \cite{martin1999active,roldan2021quantifying,tucci2022modeling}. The two-layer active model leads to a reduced-VSR of the form \eqref{eq:VSRABP} that fits the experimental data; the fitting procedure is described in Sec.\ref{S8:Fits} and Sec.\ref{sec:fit_resuts}, Supp. Mat. Some fits of the reduced-VSR are shown in Fig.~\ref{fig:FIG3}B,C, and residuals of the fits are shown in Fig.\ref{fig:S2}B-F. 

Figures \ref{fig:FIG4}D,E show $\sigma$ values obtained from OT-stretching data in the range of \tcr{trap stiffnesses $k_b=5\cdot10^{-2}-7\cdot10^{-4}$pN/nm} and OT-sensing data \tcr{with $k_b\sim 2\cdot10^{-5}$pN/nm} for healthy and ATP-depleted (passivated) RBCs. \tcr{For OT-stretching, $\sigma$ increases as $k_b$ decreases reaching $\sigma=(3\pm 1)\cdot 10^3 k_BT/s$ averaged over RBCs, for the lowest $k_b$. This value is compatible with OT-sensing measurements, $\sigma=(2\pm 1)\cdot 10^3 k_BT/s$ for healthy RBC, which is larger than for passive RBC (red and blue symbols in Figure~\ref{fig:FIG4}E).  
Moreover, $\sigma$ appears correlated with the variance of the flickering signal as measured from the position traces, ${\cal V}_x=\overline{x^2}-\overline{x}^2$ (Fig.~\ref{fig:FIG4}D). The apparent correlation demonstrates that the probe stiffness $k_b$ must be lower than the stiffness of the RBC, $k_{\rm RBC}\sim 5\cdot10^{-3}$pN/nm, to measure $\sigma$; otherwise, the active flickering of the RBC membrane is suppressed by the passive fluctuations of the bead. The correlation between $\sigma$ and ${\cal V}_x$, is also explicitly shown in Fig.~\ref{fig:S4}, where a color-map plot of the stiffness shows that we can detect active flickering and $\sigma$ only for $k_b<k_{\rm RBC}$. In fact, for the largest trap stiffness $k_b\sim 5\cdot10^{-2}$pN/nm, we get $\sigma\sim 10 k_BT/s$ (rightmost points in Fig~\ref{fig:FIG4}D), a value almost constant if the RBC is stretched up to 30pN, Fig.~\ref{fig:S4}}. The measured $\sigma$ is extensive with the bead-RBC contact area. Estimations from video images (Fig.~\ref{fig:FIG4}A and Methods) yield \tcr{circular} contact areas of $a=0.8\pm 0.2\mu m^2$ for both OT-type experiments giving the heat flux density \tcr{ $j_\sigma=\sigma/a=(3\pm 1)\cdot 10^3 k_BT/(s\cdot \mu m^2)$ for OT-stretching at low $k_{OT}$} and \tcr{$j_\sigma=(1.8\pm0.6)\cdot 10^3 k_BT/(s\cdot \mu m^2)$} for OT-sensing. \tcr{Such estimations are subject to uncertainty in the actual diameter and shape of the contact area.} Furthermore, we have analyzed the simulation data of the OT-sensing experiments based on the 3D numerical model of Ref.\cite{turlier2016equilibrium}. The active and passive trajectories for the sensing bead give \tcr{$\sigma\sim 10^4k_BT/s$ and $\sigma\sim 20k_BT/s$}, respectively (Sec.~\ref{sec:simulations}, Supp. Mat.).

For the OM experiments, we show in Fig.~\ref{fig:FIG4}C the color map of the position variance ${\cal V}_x$ (healthy, top; passive, bottom) while in Fig.~\ref{fig:FIG4}F we show the color map of $\sigma$ (circles, healthy; diamonds, passive), measured over pixels of size 50nm along the RBC contour. Both $\sigma$ and ${\cal V}_x$ reveal a RBC heterogeneous activity with average values \tcr{$\sigma=(7\pm 1)\cdot 10^3 k_BT/s$ and ${\cal V}_x=400\pm 10\,nm^2$}. Molecular maps of heterogeneous RBC deformability have been previously reported \cite{discher1994molecular}. In the active regime, $\sigma$ and ${\cal V}_x$ are anti-correlated (Pearson coefficient $\sim -0.4$) with high variance regions showing lower $\sigma$ (Fig.~\ref{fig:FIG4}G). Results for other RBCs are shown in Fig.S5. \tcr{The anticorrelation between $\sigma$ and ${\cal V}_x$ in the $\sigma$-map indicates that a larger variance does not necessarily imply a larger $\sigma$, at variance with OT-stretching (Fig.~\ref{fig:FIG4}D) and healthy versus passive data.} This counterintuitive result demonstrates the critical role of the active timescale $\tau_a$, \tcr{which, for fixed $\epsilon$}, determines the active contribution to the total variance, ${\cal V}_x={\cal V}_x^{\text{passive}}+\alpha(\tau_a)\sigma(\tau_a)$ with $\alpha(\tau_a)$ positive and monotonically increasing with $\tau_a$. It can be shown that in the high-activity limit $\tau_a\to 0$, $\sigma(\tau_a)$ saturates to a finite value whereas $\alpha(\tau_a)\sim \tau_a$, decreasing ${\cal V}_x$ (Fig.~\ref{fig:S6}). \tcr{We hypothesize that the anticorrelation observed in the $\sigma$-map derives from the highly heterogeneous $\tau_a$ (mean 0.05s and standard deviation 0.2s) but nearly constant $\epsilon$ (mean 4.4pN and standard deviation 0.2pN) across all pixel units. A constant noise amplitude $\epsilon$ with a heterogeneous $\tau_a$ suggests a uniform density of kickers but a heterogeneous ATP concentration $c_{\rm ATP}$ across the RBC surface, which modulates the ATP binding rate of the kickers,  $\tau_a^{-1}\sim k_{\rm bind}\propto c_{\rm ATP}$.}

The $\sigma$-map of a single RBC determines the finite correlation length $\xi$ for the spatially varying $\sigma$-field, a main prediction of active field theories \cite{nardini2017entropy,grandpre2021entropy} and stochastic hydrodynamics \cite{markovich2021thermodynamics}. For healthy RBCs, $\xi$ has been estimated from the spatial correlation function $C_{\sigma\sigma}(d)$, and $C_{xx}(d)$ of the traces at a curvilinear distance $d$ along the RBC contour, Fig.~\ref{fig:FIG4}H. Functions can be fitted to an exponential $\sim\exp(-d/\xi)$ with $\xi_{\sigma\sigma}\sim 0.35\pm0.05\mu$m and $\xi_{xx}\sim 0.82\pm0.02\mu$m, giving the median $\xi\sim 0.6\pm0.2\mu$m. This value is larger than the lateral resolution of the microscope (200nm). Interestingly, the structure factor of the $\sigma$-field along the cell contour shows a characteristic peak at a domain length $l\sim 1.3\mu$m, which is larger than $\xi_{\sigma\sigma}$, maybe due to the heterogeneous cortex-membrane binding-unbinding dynamics that produce differently active $\sigma$-domains (Sec.~\ref{sec:structure_factor}, Supp. Mat.). A two-layer active model in a ladder with an inter-layer coupling $k_{xx}$ further corroborates the value obtained for $\xi_{xx}$ (Sec.~\ref{sec:ladder}, Supp. Mat.). 
The average heat flux density can be estimated as \tcr{$j_\sigma=\sigma/\xi^2=(2\pm 1)\cdot 10^4 k_BT/(s\cdot \mu m^2)$} with $\xi^2$ the typical area of an entropy-producing region. In summary, for a RBC of typical surface area $A\sim 130\mu m^2$, we get: \tcr{$\sigma_{\rm RBC}=j_\sigma\cdot A=(2\pm 1)\cdot 10^5 k_BT/s$ (OT-stretching, at lowest $k_b$); $\sigma_{\rm RBC}=(2\pm 1)\cdot 10^5 k_BT/s$ (OT-sensing); and $\sigma_{\rm RBC}=(3\pm 1)\cdot 10^6 k_BT/s$ (OM)}. These values are compatible with calorimetric bulk measurements of packed RBCs, $\sigma_{\rm RBC}^{\rm bulk}=(2\pm 1)\cdot 10^6 k_BT/s$ \cite{bandmann1975clinical,backman1992microcalorimetric}, and are several orders of magnitude larger than indirect measures based on the breakdown of the fluctuation-dissipation theorem and effective temperatures \cite{turlier2016equilibrium,ben2011effective}. The significantly low $\sigma$ values obtained for passive RBCs (blue data in Figs.~\ref{fig:FIG4}E,F,G,I) validate our approach. \tcr{Our $\sigma_{\rm RBC}\sim 10^5-10^6 k_BT/s$  is higher than the values} obtained through information-theoretic measures based on the breakdown of detailed balance  \cite{roldan2021quantifying,lynn2021broken}. \tcr{Intuitively, the VSR Eqs.\eqref{eq:VSR1}, \eqref{eq:VSRABP} sets an energy balance between fluctuating positions and forces, both conjugated energy variables,} a missing feature in the thermodynamic uncertainty relation and coarse-graining models \cite{busiello2019hyperaccurate,kim2020learning,bilotto2021excess}. In general, the VSR captures most of $\sigma$ because sampling rates, 40kHz for OT-stretching, 25kHz for OT-sensing, and 2kHz for OM, are higher than the frequency of the active noise, $\sim 100s^{-1}$ for the RBC experiments (Tables \ref{tab:2}-\ref{tab:4}).\\

\noindent
{\bf\large \tcr{Discussion}}\\
\noindent
The agreement between mechanical and bulk calorimetric estimates of the RBC metabolic energy turnover suggests that the heat produced in the glycolytic pathway is tightly coupled with membrane flickering due to active kickers. Tight mechanochemical coupling is critical to an efficient free energy chemical transduction. It has been observed in processive enzymes (e.g. polymerases, transport motors, etc.) \cite{brown2019theory} and in allosteric coupling in ligand binding \cite{dokholyan2016controlling}. Tightly coupled processes are related to emergent cycles in cellular metabolism and chemical reaction networks, particularly for the relevant glycolytic cycle of RBCs \cite{wachtel2022transduction}. \tcr{A clarifying example of weak versus tight coupling is the effect of the trap stiffness in deriving $\sigma$, Fig.~\ref{fig:FIG4}D. Unless the probe stiffness is smaller than the RBC stiffness, the probe's passive fluctuations mask system's activity and $\sigma$.} Besides molecular motors and living cells, the VSR should apply to time-resolved photoacoustic calorimetry \cite{peters1988time} and enzyme catalysis, where the effective diffusion constant of the enzyme increases linearly with the heat released \cite{riedel2015heat}, a consequence of \eqref{eq:VSR1}. \tcr{Moreover, spatially resolved maps of a fraction of the full $\sigma$ for weak mechanochemical coupling provide insight into the structural features underlying heat dissipation in biological cells. In a wider context, the VSR applies to nonlinear systems, from non-Gaussian active noise to non-linear potentials, Sec.\ref{sec:nonlinear}.}
Finally, we stress that different models can fit the experimental data. However, the power of the VSRs, \eqref{eq:VSR1} and \eqref{eq:VSRABP}, is given by the constraint imposed by the sum of variances over the experimental timescales. \tcr{By fitting the experimental data to a single function, the total variance $V_T(t)$ over several decades, the contribution of dissipative processes over multiple timescales are appropriately weighted in the sum balance.} This distinguishes our approach from plain model fitting of the experimental power spectrum to derive the model parameters \cite{tucci2022modeling} that may lead to inaccurate estimations (Sec.~\ref{sec:VSRvsPS}, Supp. Mat.). In this regard, the VSR links modeling with energetics.

\let\oldaddcontentsline\addcontentsline
\renewcommand{\addcontentsline}[3]{}

\begin{acknowledgments}
\noindent
\textbf{Funding}:
M.G. and F.R. are supported by the Spanish Research Council Grant  PID2019-111148GB-100. D.H. and F.M. are supported by the Spanish Research Council Grant  PID2019-108391RB-100. T.B. is supported by the European Research Council (consolidator grant 771201).
M.B. is supported by the research grant BAIE$\_$BIRD2021$\_$01 of the University of Padova.
F.R. is supported by ICREA Academia 2018.
\textbf{Author contributions}: 
I.D., M.B., and F.R. conceptualized;
M.G., D.H., T.B., and F.M. collected and cured the data.
I.D. wrote the software for data analysis and took care of visualization.
I.D. and M.G. analyzed the data.
F.R. administered the project.
I.D., M.B., and F.R. wrote the original draft.
All the authors discussed the results and implications of the methodology and commented on the manuscript.
\textbf{Competing interests}:
The authors declare no competing financial interests.
\textbf{Data and materials availability}:
All data needed to evaluate the conclusions in the paper and the code for fitting the VSR are available \tcr{at Dryad \cite{VSR_data_code}}.  Figures 1a, 2a, S1, and inset of 3a have been created with BioRender.com.

\end{acknowledgments}

\nocite{peliti2021stochastic,seifert2019stochastic,sekimoto1998langevin,gin16,macieszczak2018unified,horowitz2020thermodynamic,falasco2020unifying,dit19,paoluzzi2022scaling,cates2022stochastic}

\bibliographystyle{naturemag}
\bibliography{BibNonEq}

\let\addcontentsline\oldaddcontentsline

\clearpage
\onecolumngrid
\renewcommand{\thefigure}{S\arabic{figure}}
\renewcommand\theequation{S\arabic{equation}}
\renewcommand{\thesection}{S\arabic{section}}
\setcounter{figure}{0}
\setcounter{equation}{0}

\setcounter{page}{1}


\hspace{1cm}
\begin{center}
{\LARGE Supplementary Material for}\\
\vspace{0.4cm}
\textbf{\Large Variance Sum Rule for Entropy Production}\\
\hspace{0.5cm}

{I. Di Terlizzi$^{*}$, M. Gironella$^{*}$, D. Herraez-Aguilar, T. Betz, F. Monroy, M. Baiesi, F. Ritort}

{\small Correspondence to: ritort@ub.edu}
\end{center}
\vspace{1cm}

\noindent

\noindent
\textbf{The PDF file includes:}
\begin{itemize}
    \setlength\itemsep{-0.4em}
    \item[] Materials and Methods
    \item[] Supplementary Text
    \item[] Figs. S1 to S15
    \item[] Tables S1 to \tcr{S5} 
\end{itemize}


{
\tableofcontents
}

\clearpage


\clearpage

\section*{Materials and Methods}\label{S0:Methods} 
\textbf{Optical trap experiments with beads.} 
Experiments with colloidal particles (Figs.\ref{fig:FIG1},\ref{fig:FIG2}) were done in a miniaturized version of an optical tweezers instrument described in \cite{dieterich2015single}. Measurements in Figures \ref{fig:FIG1} and \ref{fig:FIG2} were performed with highly stable miniaturized laser tweezers in the dual-trap mode. The instrument directly measures forces by linear momentum conservation. In all experiments, we used polystyrene calibration beads of 3$\mu$m diameter.  Piezo actuators bend the optical fibers and allow us to move the trap while measuring the trap position using a light lever that deflects a fraction of the laser beam to a position-sensitive detector (PSD). Force and trap position measurements are acquired at 100 kHz bandwidth using a data acquisition board (PXI-1033, National Instruments, Austin, TX).

\textbf{OT-stretching RBC experiments.} For the RBC experiments, human RBCs were obtained by finger pricking of a healthy donor. The PBS solution contains 130 mM NaCl, 20 mM K/Na phosphate buffer, 10 mM glucose, and
1 mg/mL BSA. For the experiments, 4$\mu$L of blood was diluted in 1mL of PBS. The OT-stretching consists of three steps: 1) the RBC is non-specifically attached to a bead that is captured in an optical trap of stiffness 56pN/$\mu$m while the RBC remains outside the optical trap; 2) the bead is brought to the tip of a micropipette where it remains immobilized while the RBC remains attached on the other side of the bead; (3) a second bead is captured by the optical trap and brought to the opposite end of the RBC to form a dumbbell configuration. All measurements were made at 40kHz with a data acquisition board (PXI-1033, National Instruments, Austin, TX). Bead-RBC contact areas (Fig.~\ref{fig:FIG4}A) were estimated using a multiscale feature extractor based on a Gaussian pyramid representation of the raw image followed by a Laplacian reconstruction. 

\textbf{OT-sensing RBC experiments.}  Experiments have been performed as described previously \cite{turlier2016equilibrium}. RBCs were obtained from a healthy donor by finger pricking. After washing in PBS-based cell medium (CM), cells were biotinylated using a 0.5mM NHS-PEG3400-biotin (Nektar Therapeutics, San Carlos, CA) solution in CM. Streptavidin-coated, 3.28µm diameter polystyrene beads were incubated with the biotinylated RBC, and four beads were attached to the RBC using a time-shared optical tweezers system. For the OT-sensing, three of the four beads were trapped (trap stiffness: 1.2pN/$\mu$m), and the cell and bead system was moved 20$\mu$m away from the surface to avoid any unspecific substrate interaction. To detect the free fluctuations of the fourth bead (probe bead), the laser power on this bead was reduced to 0.1 mW which is insufficient to trap the bead (trap stiffness $<$ 24fN/$\mu$m). The particle motion was recorded using a position-sensitive detector (PSD) placed in a plane conjugate to the back focal plane of the light-collecting objective (Obj2 in Fig.\ref{fig:FIG4}B). Low-frequency noise peaks known to originate from the stage were filtered in Fourier space. For OT-sensing, bead-RBC contact areas were estimated as described in \cite{turlier2016equilibrium}. 

\textbf{OM RBC experiments.} Human RBCs were extracted from the blood of healthy donors and then separated by centrifugation at 5000g for 10 min at 4ºC, rinsed with PBS, and, finally, diluted (1:15) with PBS supplemented with 10 mM glucose, and 1 mg/mL bovine serum albumin \cite{rodriguez2015direct}. High-resolution time-lapse Optical Microscopy (OM) was performed using a phase contrast inverted microscope (NikonEclipse2000Ti) armed with a 100 W TI-12 DH Pillar Illuminator, an LWD 0.52 collimator, and a 100x oil immersion objective (PlanApoVC, N.A. 1.4; Nikon). RBC flickering was captured at the equatorial plane with a FASTCAM SA3 camera (Photron), with an effective pixel size of 50x50 nm$^2$. Movies were recorded for 10s, with a sampling frequency of 2000 frames per second (2kHz).

\section{Variance Sum Rule}\label{S1:VSRDer}
We derive the variance sum rule (VSR) and Eqs.(2-4) in the main text for Markovian stochastic diffusion dynamics for one degree of freedom in one dimension in contact with a thermal bath at temperature $T$. Moreover, the results can be readily extended to an arbitrary number of degrees of freedom, dimensions, and reservoirs at different temperatures. 

The overdamped Langevin equation describes the dynamics,
\begin{equation}\label{eq:MultiLE}
\dot{x}_t = \mu \, F_{t} + \sqrt{2\,D}~ \eta_t  \,,
\end{equation}
where the white noise $\eta$ is Gaussian with first two moments $\mean{\eta_t}= 0$ and $\mean{\eta_t \eta_{t'}} = \delta(t-t')$ (52). The diffusion constant equals $D = k_{B}T \mu$, with mobility $\mu$ and temperature $T$.

 For simplicity, we consider a non-conservative force that derives from a time-dependent potential, $F_{t}=-\partial_x U(x_t-\lambda_t)$ with $\lambda_t$ an explicit time-dependent external protocol and $U(x)$ a potential energy function. Here, we focus on a NESS that is homogeneous in time, excluding the mathematically more complicated periodic NESS. A paradigmatic example of a NESS is a system that moves in a lab frame $\lambda_t = vt$ at a constant velocity $v$. In this case, it is helpful to consider the co-moving frame and change variables $x_t \to y_t=x_t - vt$. This setting is essential for the example of the optically trapped bead dragged through water discussed in Figure 1 (main text).  
 
 In a NESS, the correlation functions between observables $A,B$ that depend on time via the coordinate $x_t$, $A_{t}\equiv A(x_t)$ and $B_{t}\equiv B(x_t)$. We define the connected correlation function in the NESS,  
 \begin{equation}
     C_{\!AB}(t) =  \mean{A_{t}B_{0}} - \mean{A_{0}} \, \mean{B_{0}}
 \end{equation}
where the time 0 stands for an arbitrary initial time in the NESS and $\mean{A_{t}}=\mean{A_{0}}$ (same for $B$) are time-independent. In the following, we will also need time derivatives of the correlation functions, denoted as 
 $ \overset{\,\bm .}{C}\,{\vphantom{C}}_{\!AB}(t) \equiv \partial_t C_{\!AB}(t)$. Their limit
$\overset{\,\bm .}{C}\,{\vphantom{C}}_{\!AB}(0^+) = \lim_{t\to 0^+} \overset{\,\bm .}{C}\,{\vphantom{C}}_{\!AB}(t)$ is always taken for positive $t$ approaching zero. 
 
\subsection{Derivation}
We start by integrating equation \eqref{eq:MultiLE} from 0 to $t$  and rearranging terms,
\begin{equation}
    x_t-x_0 -  \int_{0}^{t}\mathrm{d}t'  \mu~\!F_{t'} = \sqrt{2~\!D}\int_{0}^{t} \mathrm{d}t'   \eta_{t'}  \,,
    \label{eq:deltax}    
\end{equation}
By taking the variance of both sides, we get
\begin{equation}\label{Sum_rule_step2}
{\cal V}_{\Delta x}(t) +\mu^2{\cal V}_{\Sigma_ F}(t) = 2 D t +2 \mu \int_{0}^{t}\mathrm{d}t' \left(C_{xF}(t')-C_{Fx}(t') \right) \,.
\end{equation}
where ${\cal V}_{\Delta x}(t)$ denotes the variance of $\Delta x_t=x_t-x_0$ and ${\cal V}_{\Sigma_F}(t)$ is the variance of $\Sigma_F(t)=\int_{0}^{t}\mathrm{d}t'\!F_{t'}$. For the last term, we used the properties of the Gaussian white noise. The excess variance as defined in the main text Eq.\eqref{eq:S1} gives the VSR in Eq.\eqref{eq:VSR1},
\begin{equation}\label{Sum_rule_supp}
{\cal V}_{\Delta x}(t) +\mu^2{\cal V}_{\Sigma_ F}(t) = 2 D t +{\cal S}(t) \,.
\end{equation}

\subsection{$d$-dimensional VSR}
The VSR can be generalized to $d$ dimensions. Each degree of freedom $x^i$, with $1 \le i\le d$, evolves with the overdamped Langevin equation
\begin{equation}
\label{eq:LE}
\begin{split}
     x^i_t &
     =  \sum_{j=1}^d 
     \left[
     \mu^{ij}\, F^j_{t} + \sqrt{2\,D^{ij}} \cdot \xi^j_t
     \right]
\end{split}
\end{equation}
where a constant symmetric molbility matrix $\mu^{ij}$ multiplies the components $F^i$ of the force, the diffusion matrix is equal to $D^{ij}=k_B T\,\mu^{ij}$ and $\xi^{i}_{t'}$ is a $d$-dimensional Gaussian white noise ($\langle \xi^{i}_{t'} \rangle = 0$ and $\langle \xi^{i}_{t'} \xi^{i}_{t''} \rangle = \delta_{ij}\, \delta(t'-t'')$).
We express covariances as a function of connected correlation functions between variables $A^i_{t'}$ and $B^j_{t''}$, which are homogeneous in time $t=t'-t''$:
 \begin{equation}\label{CorrF}
     C_{\!AB}^{ij}(t) = C_{\!AB}^{ij}(t',t'') = \mean{A^i_{t'}B^j_{t''}} - \mean{A^i_{t'}}\,\mean{B^j_{t''}} \, .
 \end{equation}
Thus, by defining the covariances
\begin{align}
    \mathcal{V}^{ij}_{\Delta x}(t) 
    &= \mean{ {\Delta x}^i_{t} {\Delta x}^j_{t} }
    - \mean{{\Delta x}^i_{t}}\,\mean{{\Delta x}^j_{t}}\, ,
    \\
    \mathcal{V}^{ij}_{\Sigma_F} (t)
    &= \int_{0}^t\mathrm{d}t'\int_{0}^t\mathrm{d}t''
    C^{ij}_{F F}(t',t'')
    \,,
\end{align}
the $d$-dimensional VSR for two degrees of freedom $i,j$ becomes
\begin{equation}
\label{eq:VSR_multi}
{\cal V}^{ij}_{\Delta x}(t) +
 \sum_{k=1}^d  \sum_{l=1}^d  \mu^{il} \mu^{jk} {\cal V}^{lk}_{\Sigma_ F}(t) = 2 D^{ij}\, t + \mathcal{S}^{ij}(t)\,,
\end{equation}
in which the excess (co)variance
\begin{equation}
\label{eq:S} 
\mathcal{S}^{ij}(t) = 2 \int_{0}^{t}\mathrm{d}t'\, \sum_{k=1}^d \mu^{jk}\left(C^{ik}_{xF}(t')-C^{ki}_{Fx}(t') \right)^S \, .
\end{equation}
includes a symmetrized correlation function $(C^{ik})^{S} \equiv (C^{ik}+C^{ki})/2$.  
In equilibrium, correlation functions are time-symmetric, and each $\mathcal{S}^{ij}(t)$ vanishes.

\section{Entropy production rate}
\label{SS2:EPfromVSR}
The entropy production rate $\sigma$ is a key quantity in stochastic thermodynamics (52,53) and is constant in a NESS. Hence, the entropy $\sigma\,dt$ produced in a time interval $dt$ is constant. According to stochastic energetics (54), for a system described by \eqref{eq:MultiLE}, Sekimoto has introduced a formula for $\sigma$,
\begin{equation}
\sigma\,dt = \frac{1}{k_B T}\mean{F_{t}\circ {dx}_{t}}
\label{eq:EP_sekimoto}    
\end{equation}
where we have expressed $\sigma$ in units of $k_BT$ and $\circ$ denotes a Stratonovich product.
The average Stratonovich product of force and an infinitesimal displacement gives the amount of heat $\mean{dQ} = \mean{F_{t}\circ d{x}_{t}}$ delivered to the environment in a time $dt$, 
\begin{equation}\label{eq:sek_der}
\begin{split}
    \mean{F_{t}\circ d{x}_{t}} 
    &= \frac 1 2 \mean{\left(F_{t+dt}+F_{t}\right)\left(x_{t+dt}-x_{t}\right)}\\
    &= \frac 1 2 \mean{\left(F_{t+dt}+F_{t}\right)\left(y_{t+dt}-y_{t} + v dt\right)}\\
    &= \frac 1 2 \left[ C_{yF}(dt)-C_{Fy}(dt)\right] +  \frac{v^2}{\mu}dt
\end{split}
\end{equation}
where $y_t=x_t-vt$ in the general case where there is mean velocity $v$. Between the second and third row we used $\mu\mean{F_t}=v$, $\mean{x_t-x_{0}}=v t$ obtained by averaging Eqs.\eqref{eq:MultiLE} and  \eqref{eq:deltax}. We have also used the fact that $\mean{y_t},\mean{F_t},\mean{y_tF_t}$ are time independent. By performing a Taylor expansion up to order $dt$, one gets
\begin{equation}\label{eq:sek_der_3}
\begin{split}
    \mean{F_{t}\circ d{x}_{t}} 
    & = \frac 1 2 \left[ \overset{\,\bm .}{C}\,{\vphantom{C}}_{\!yF}(0^+)-\overset{\,\bm .}{C}\,{\vphantom{C}}_{\!Fy}(0^+)\right] dt +  
    \frac{v^2}{\mu}\,dt \\
    & = \frac 1 2 \left[ \overset{\,\bm .}{C}\,{\vphantom{C}}_{\!xF}(0^+)-\overset{\,\bm .}{C}\,{\vphantom{C}}_{\!Fx}(0^+)\right] dt +  
    \frac{v^2}{\mu}\,dt \, .
\end{split}
\end{equation}
With \eqref{eq:sek_der_3} we rewrite \eqref{eq:EP_sekimoto} as
\begin{equation}
\sigma =   
    \frac{v^2}{D} +\frac{1}{2k_BT} \left[ \overset{\,\bm .}{C}\,{\vphantom{C}}_{\!xF}(0^+)-\overset{\,\bm .}{C}\,{\vphantom{C}}_{\!Fx}(0^+)\right] 
\label{eq:EP_sekimoto_final}    
\end{equation}
with $D=k_BT\mu$. From the excess variance Eq.\eqref{eq:S1} we get,
\begin{equation}
\sigma =   
    \frac{v^2}{D} +\frac{1}{4D} \partial_{t}^2 {\cal S}|_{t=0}
\label{eq:sigma2}    
\end{equation}
which is Eq.\eqref{eq:EP0} in the main text. Here it is expressed in rate units (e.g., $1/s$) whereas, in the main text, $\sigma$ is expressed in power units (e.g., $k_BT/s$). 

\subsection{Entropy production from the VSR}
 Consider now the VSR \eqref{Sum_rule_step2} and divide both sides by $D$,
 \begin{equation}
 \label{Sum_rule_entropy}
\frac{1}{k_B T}\left(\frac{1}{\mu}{\cal V}_{\Delta x}(t) +\mu {\cal V}_{\Sigma_ F}(t)\right) = 2t + \frac{2}{k_B T}  \int_{0}^{t}\mathrm{d}t' \left(C_{xF}(t')-C_{Fx}(t') \right) \, .
\end{equation}
By taking the second order derivative evaluated at $t=0$ of \eqref{Sum_rule_entropy} and using it along with \eqref{eq:EP_sekimoto_final}, one readily sees that 
\begin{equation}
    \sigma = \frac{v^2}{D}+\frac{1}{k_B T}\left(\frac{1}{4\mu}{\partial^2_t\cal V}_{\Delta x}(t)|_{t=0} +\frac{\mu}{2} {\cal V}_{F}(0) \right)\; ,
\end{equation}
because $\partial^2_t {\cal V}_{\Sigma_ F}(t)|_{t=0}=2 \mathrm{Var}(F_{0},F_{0})\equiv 2  {\cal V}_{F}(0) $ for reasons of symmetry.  For the experimental data presented in this study, and to improve convergence, we have calculated the second derivative term $\partial^2_t{\cal V}_{\Delta x}(t)|_{t=0}$ by fitting the variance ${\cal V}_{\Delta x}(t)$ to a sum of time-dependent exponential functions.

\subsection{Entropy production from the $d$-dimensional VSR}
In $d$ dimensions, contributions to the entropy production rate come from the curvature of each $\mathcal{S}^{ij}(t)$ for $t\to 0$,
\begin{equation}
\label{eq:sigma-d}
\sigma = \frac{1}{k_B T}
\sum_{i,j} 
\left[
v^{i}\, (\mu^{-1})^{ij}\,v^{j}+
\left.{(\mu^{-1})^{ij} \partial^2_t\mathcal{S}}^{ij}(t)\right|_{t=0} 
\right] \, ,
\end{equation}
where $v^i = \mean{\dot{x}^i}$ are the components of the stationary mean velocity.
We may rewrite \eqref{eq:sigma-d} as
\begin{equation}
\label{eq:sigma_VSR}
\sigma = \frac{1}{k_B T} \sum_{i,j} 
\left[
v^{i}\, (\mu^{-1})^{ij}\,v^{j} +
\left.\frac{1}{4}{(\mu^{-1})^{ij}\partial^2_t\cal V}^{ij}_{\Delta x}(t)\right|_{t=0} +\frac{1}{2}\mu^{ij} {\cal V}^{ij}_{F}
\right]\, ,
\end{equation}
where ${\cal V}^{ij}_{F} = \mean{F^i F^j} - \mean{F^i}\,\mean{F^j} $ is the covariance between forces $F^i$ and $F^j$.

\section{Optically trapped bead dragged through water}\label{S3:ExDragging}
A colloidal particle is confined by an optical trap in water, which moves at a uniform speed $v$. The corresponding potential energy is $U(x,t) = \frac k 2 (x-v t)^2$, and the force on the particle is $F_t = k\cdot(vt-x)$.

The position $y_t=x_t-vt$ of the particle in the co-moving frame follows the overdamped diffusion equation 
\begin{align}
    \gamma \dot{y}_t=-k y_t-\gamma v+\sqrt{2 k_B T\gamma} \eta_t
\end{align}
where  $\eta$ is a Gaussian white noise, $\overline{\eta_t\eta_s}= \delta(t-s)$. Here, $\gamma=\mu^{-1}=6\pi\nu R$ is the friction coefficient (with $R$ the bead radius and $\nu$ the shear viscosity), leading to a relaxation time $\tau_r=\gamma/k$ in the trap. Analytical calculations give 
\begin{align}
    {\cal V}_{\Delta x}(t)
    &=
    {\cal V}_{\Delta y}(t)=
    \frac{2k_BT}k (1-e^{-t/\tau_r})
    \\
    {\cal V}_{\Sigma_ F}(t)
    &=
    2k_BT\gamma \left[t -\tau_r(1-e^{-t/\tau_r})\right]
\end{align}
Summing the two variances, we get ${\cal V}_T(t)={\cal V}_{\Delta x}(t)+\mu^2{\cal V}_{\Sigma_ F}(t)=2Dt$ and $\mathcal{S}(t)=0$ from Eq.(\ref{eq:VSR1}). Therefore, as expected, the entropy production rate reduces to $\sigma=v^2/\mu=\gamma v^2$. For a bead dragged by a generic confining potential $U(x-v t)$, one still has ${\cal S}(t)=0$ and $\sigma=\gamma v^2$.

\section{Reversed thermodynamic uncertainty relation}\label{S4:RTUR}
In this section, we show how to obtain a reversed version of the thermodynamic uncertainty relation (TUR) (16,55,56,57,58) for systems dragged at a constant velocity and for two observables: the displacement $x_t-x_0$ and the thermodynamic work $W_t$.
Indeed, for generic confining potentials $U(x-v t)$, Eq.(\ref{eq:VSR1}) gives two inequalities: ${\cal V}_{\Delta x}(t)\le 2Dt$ and ${\cal V}_{\Sigma_ F}(t)\le (2D/\mu^2)t$. The first inequality states that a particle's diffusion cannot be larger than $2Dt$. This does not hold for superdiffusive systems where ${\cal S}\ne 0$. The second inequality can be rewritten in terms of the work exerted on the bead by the optical trap, $W_t=v\Sigma_F(t)=v\int_0^tdsF_s$, and therefore ${\cal V}_{W}(t)/(v^2\gamma^2)\le 2Dt$. From $\overline{W_t}=\gamma v^2 t=\sigma t$, we get a reversed thermodynamic uncertainty relation and an upper bound for $\sigma$ (Inset of Fig.\ref{fig:FIG1}C, main text).
\begin{equation}
   \frac{\sigma}{k_BT}\le  \frac{2\overline{W_t}^2}{t\,{\cal V}_{W}(t)}\,.
   \label{eq:RTURb}
\end{equation}

\section{Stochastic switching trap (SST)}
\label{S5:SST}

We consider a Brownian particle at a temperature $T$ and with mobility $\mu$ in a harmonic trap whose center $\lambda_t$ jumps stochastically between the positions $\lambda_- = 0$ and $\lambda_+ = \Delta\lambda$, i.e. its potential energy is $U(x_t,\theta_t) = k \left(x_{t}-\theta_{t} \Delta\lambda \right)^{2}/2$, with dichotomic stochastic variable $\theta_t = \{0,1\}$ that is uncorrelated with the noise $\eta_t$ and undergoes a Markovian jumping dynamics with jumping rates $w_-$ for the $\lambda_-\to \lambda_+$ transition and $w_+$ for the reverse one. 
By defining $w = w_-+w_+$, the stationary average of $\theta_{t}$ can be written as $q = \langle \theta_{t} \rangle =w_-/w$. A stochastic diffusion equation gives the dynamics of the Brownian particle,
\begin{equation}\label{LE_switch_present}
     \dot{x}_t = -\mu k \left( x_t - \Delta\lambda \theta_t\right)+ \sqrt{2\,k_{B}T\mu}~ \eta_t\,.
\end{equation}
In the NESS $\langle \dot x_{t} \rangle = 0$. Hence, the stationary average of the particle position is $\langle x_{t} \rangle= q\,\Delta \lambda$. Like in the previous section \ref{S3:ExDragging}, the relaxation rate of the bead in the trap is given by $w_r = 1/\tau_r=\mu k$. 

As we will show later, all the quantities we are interested in can be calculated in terms of the stationary correlation functions $C_{xx}(t) = \langle x_{t}\,x_{0} \rangle$, $C_{x\theta}(t) = \langle x_{t}\,\theta_{0} \rangle$, $C_{\theta x}(t) = \langle \theta_{t}\,x_{0} \rangle$ and $C_{\theta\theta}(t) = \langle x_{t}\,\theta_{0} \rangle$ for $t \ge 0$. To compute these correlations, we turn to a fine time-step description of the dynamics,
\begin{subequations}
\label{LE_switch_disc}
\begin{align}
x_{t+dt} &= x_t -w_r\,x_t\,dt +\mu\,k\,\Delta\lambda\,\theta_t\,dt + \sqrt{ 2\,k_{B}T\mu} \,d\mathcal{B}^{x}_t
\label{LE_switch_disc_a}
\\
\theta_{t+dt} &= \theta_t +(1-2\,\theta_t)\Theta_H(w_{\theta_{t}}dt -r)
\label{LE_switch_disc_b}
\end{align}
\end{subequations}
where $r$ is random variable with uniform probability distribution on $[0,1]$, $\Theta_H(\cdot)$ is the Heaviside step function, $w_{\theta_{t}}$ represents the jumping rate at the state corresponding to $\theta_t$, and $\,d\mathcal{B}^{x}_t$ is the integral of white noise over $dt$. By multiplying \eqref{LE_switch_disc_a} respectively by $x_0$ or $\theta_0$ and then taking stationary averages, one obtains first-order differential equations that read
\begin{subequations}
\label{eq:dCss}
\begin{align}
\label{corr_1_ch6}
    \partial_{t} C_{xx}(t) 
    &= - w_r\,C_{xx}(t) + \mu\, k\,\Delta\lambda \,C_{\theta x}(t) \,,
    \\
    \label{corr_2_ch6}
    \partial_{t} C_{x \theta}(t) 
    &= - w_r\,C_{x\theta}(t) + \mu\,k\,\Delta\lambda \,C_{\theta \theta}(t)
      \,.
\end{align}
In a similar way, by multiplying \eqref{LE_switch_disc_b} by $x_0$ or $\theta_0$, one can show that the following equations hold, 
\begin{align}\label{corr_3_ch6}
    \partial_{t} C_{\theta x}(t) 
    &= w_-\langle x_{t} \rangle - w \,C_{\theta x}(t) \,,
    \\
    \label{corr_4_ch6}
    \partial_{t} C_{\theta \theta}(t) 
    &= w_-\,q -w\,C_{\theta \theta}(t) \,.
\end{align}
\end{subequations}
We solve these linear equations in terms of the correlation functions at time $t=0$ calculated from \eqref{LE_switch_disc}:
\begin{equation}
    \begin{split}
    C_{xx}(0) 
    &= \Delta\lambda^2 q^2  + \frac{k_B \, T}{k}+\frac{k\,\Delta\lambda^2 \,q(1-q)\,\mu}{(w+w_r)}\\
    C_{\theta x}(0) 
    &= \Delta\lambda\, q^2 + \frac{k\,\Delta\lambda\, \,q(1-q)\mu}{w+w_r} \\
    C_{x\theta}(0) &=C_{\theta x}(0) \\
    C_{\theta\theta}(0) &= q \,.
\end{split}
\end{equation}
With these quantities we solve equations \eqref{eq:dCss},  finding
\begin{equation}
    \begin{split}
    C_{xx}(t) = & \Delta\lambda^2 q^2 + \left(\frac{k_B \, T}{k}+\frac{k\,\Delta\lambda^2 q(1-q)\,\mu}{(w+w_r)}\right)\text{e}^{-w_r\,t} + \frac{(k\,\Delta\lambda)^2q(1-q)\,\mu^{2}}{w^{2}-\mu^{2}\,k^{2}} \left(\text{e}^{-w_r\,t}-\text{e}^{-w\,t} \right) \\[6pt]
    C_{\theta x}(t)  =& \Delta\lambda\, q^2 + \frac{k\,\Delta\lambda \,q(1-q)\mu}{w+w_r} \text{e}^{-w_r\,t} \\
    C_{x\theta}(t) =& \Delta\lambda\, q^2 + \frac{k\,\Delta\lambda \,q(1-q)\mu}{w+w_r} \text{e}^{-w_r\,t} + \frac{k\,\Delta\lambda \,q(1-q)\mu}{w-w_r} \left(\text{e}^{-w_r\,t}-\text{e}^{-w\,t} \right)  \\
    C_{\theta\theta}(t) = &q^{2} + q(1-q)\text{e}^{-w\,t} \, .
\end{split}
\end{equation}
Introducing $\epsilon=k\,\Delta\lambda\sqrt{q(1-q)}$, we compute the variance of the relative displacement, 
\begin{equation}\label{switch_var_pos}
\begin{split}
    {\cal V}_{\Delta x}(t) 
    = 2\left[\left(\frac{k_B \, T}{k}+\frac{\epsilon^2\,\mu}{k(w+w_r)}\right)(1-\text{e}^{-w_r\,t}) 
    + \frac{\epsilon^2\,\mu^{2}}{w^{2}-w_{r}^{2}} \left(\text{e}^{-w\,t}-\text{e}^{-w_r\,t} \right)\right]\, .
\end{split}
\end{equation}
The variance of the time-integrated force  $F_t=k\cdot(\Delta \lambda\, \theta_t -x_t)$ is 
\begin{equation}\label{switch_var_sum_forc}
\begin{split}
    {\cal V}_{\Sigma_F}(t) = \frac{2\,k_{B}T\,t}{\mu}+ \frac{2\,k_{B}T}{\mu^2\,k} \left(1- \text{e}^{-w_r\,t}\right)+\frac{2\,\epsilon^2}{w_r(w+w_r)}\left(1 - \frac{w\,\text{e}^{-w_r\,t}}{w-w_r}+ \frac{w_r\,\text{e}^{-w\,t}}{w-w_r} \right) \, .
\end{split}
\end{equation}
Finally, we compute ${\cal S}(t)$ by means of the VSR and equations \eqref{switch_var_pos} and \eqref{switch_var_sum_forc}, 
\begin{equation}\label{viol_fac_switch}
 {\cal S}(t) = 
 \frac{4~\!\epsilon^2}{k(w+w_r)}
 \left(1 - \frac{w~\!\text{e}^{-w_r~\!t}}{w-w_r}+ \frac{w_r~\!\text{e}^{-w~\!t}}{w-w_r} \right)\,.
\end{equation}
From Eq.\eqref{eq:sigma2}, one finds $\sigma$ for the SST expressed in power units,
\begin{equation}
\sigma_{\rm th} = \epsilon^2\mu \frac{ w}{w+w_r}\,.
\end{equation}

\section{Reduced VSR}\label{S6:RedVSR}
When dealing with hidden degrees of freedom, a reduced form of the VSR expressed in terms of $x_t$ only is needed. In particular, we consider the case of an experimental device (optical tweezers, AFM, etc.) where the measured force is linear with the probe's displacement, $F_t^M=-kx_t$, where $k$ is the stiffness of the device. The total force acting on the probe is given by $F_t=F_t^M+F_t^I$ where $F_t^I$ cannot be measured. Equation \ref{eq:MultiLE} can be written as,
\begin{equation}
    \dot x_t +k\mu x_t = \mu F_t^I+ \sqrt{2D}\eta_t  \, .
    \label{eq:redVSR_step1}
\end{equation}
By integrating both sides from $0$ to $t$ and taking the variance as in Eqs.\eqref{eq:deltax}, \eqref{Sum_rule_step2}, one gets a reduced form of the VSR (reduced-VSR),
\begin{equation}
   {\cal V}_{\Delta x}(t)+\mu^2k^2\,{\cal V}_{\Sigma_x}(t)=2Dt+{\cal \tilde{S}}(t) \,,
\label{eq:VSRABP_supp}
\end{equation}
where ${\cal V}_{\Sigma_x}(t)=\overline{\Sigma^2_x(t)}-\overline{\Sigma_x(t)}^2$, $\Sigma_x (t)=\int^t_0ds\, x_s$ and 
\tcr{\begin{equation}
    {\cal \tilde{S}}(t) = 2\mu^2\int_0^t
ds \int_0^s du \,C_{F^I F^I}(u) +2\mu\sqrt{2D}\int_0^t
ds \int_0^s du \,C_{F^I \eta}(u) \,
\label{eq:reducedSstep1}.
\end{equation}}
Let us consider the case where $F^I_t$ does not depend on $x_t$, therefore $C_{F^I \eta}(t)=0$. An example is the active Brownian particle (ABP) defined in Eq.\eqref{eq:ABP} in the main text where $F^I_t = f^a_t$ stands for an active force. We get the following averages: $\overline{f_t^af_s^a}=\epsilon^2 e^{-|t-s|/\tau_a}$, $\overline{f_t^a}=0$, $\overline{f_t^a \eta_s}=0$, and Eq.\eqref{eq:reducedSstep1}
\begin{equation}\label{eq:VSR_ABP_supp}
    {\cal \tilde{S}}^{ABP}(t) = 2\mu^2\int_0^t
ds \int_0^s du\, \overline{f_u^af_0^a}=2 \epsilon^2\mu^2\tau_a[t-\tau_a(1-e^{-t/\tau_a})]\, .
\end{equation}
In general, one can further show that the reduced excess variance ${\cal \tilde{S}}(t)$ in Eq.\eqref{eq:reducedSstep1} can be rewritten as
\tcr{\begin{equation}\label{eq:redS_comp}
    {\cal \tilde{S}}(t) = 2\mu\int_0^t
ds \int_0^s du \,\left(\mu k C_{F^I x}(u)- \overset{\,\bm .}{C}\,{\vphantom{C}}_{\!F^I x}(u)\right)  \, .
\end{equation}}
The full expression in Eq.\eqref{eq:redS_comp} for ${\cal \tilde{S}}(t)$ will be needed for the RBC model discussed in the next section \ref{S7:RBCmodel}.

\section{RBC model}\label{S7:RBCmodel}
\begin{figure}
   \centering
   \includegraphics[width=17cm]{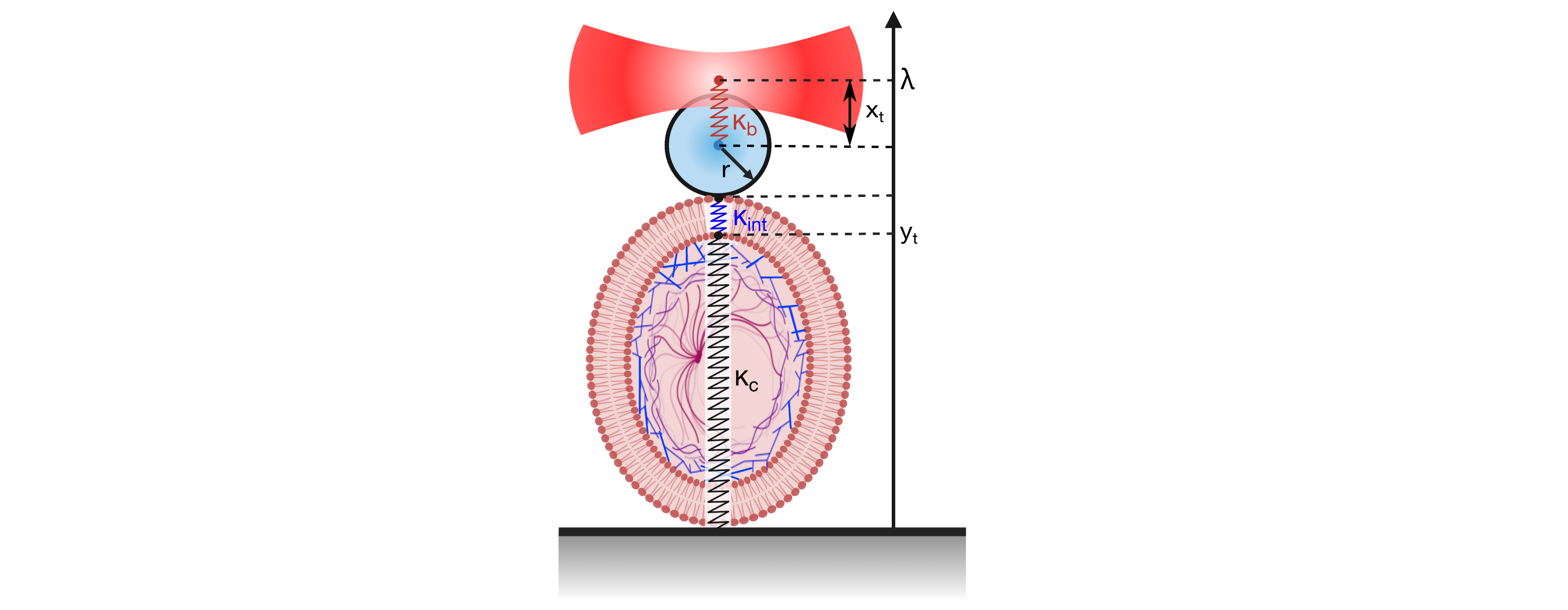} 
   \caption{\textbf{RBC model.} Sketch of RBC structure in terms of linear springs leading to Eq. \eqref{eq:ABPRBC}.} 
   \label{fig:S1_RBC}
\end{figure}

\begin{figure}
   \centering
   \includegraphics[width=\linewidth]{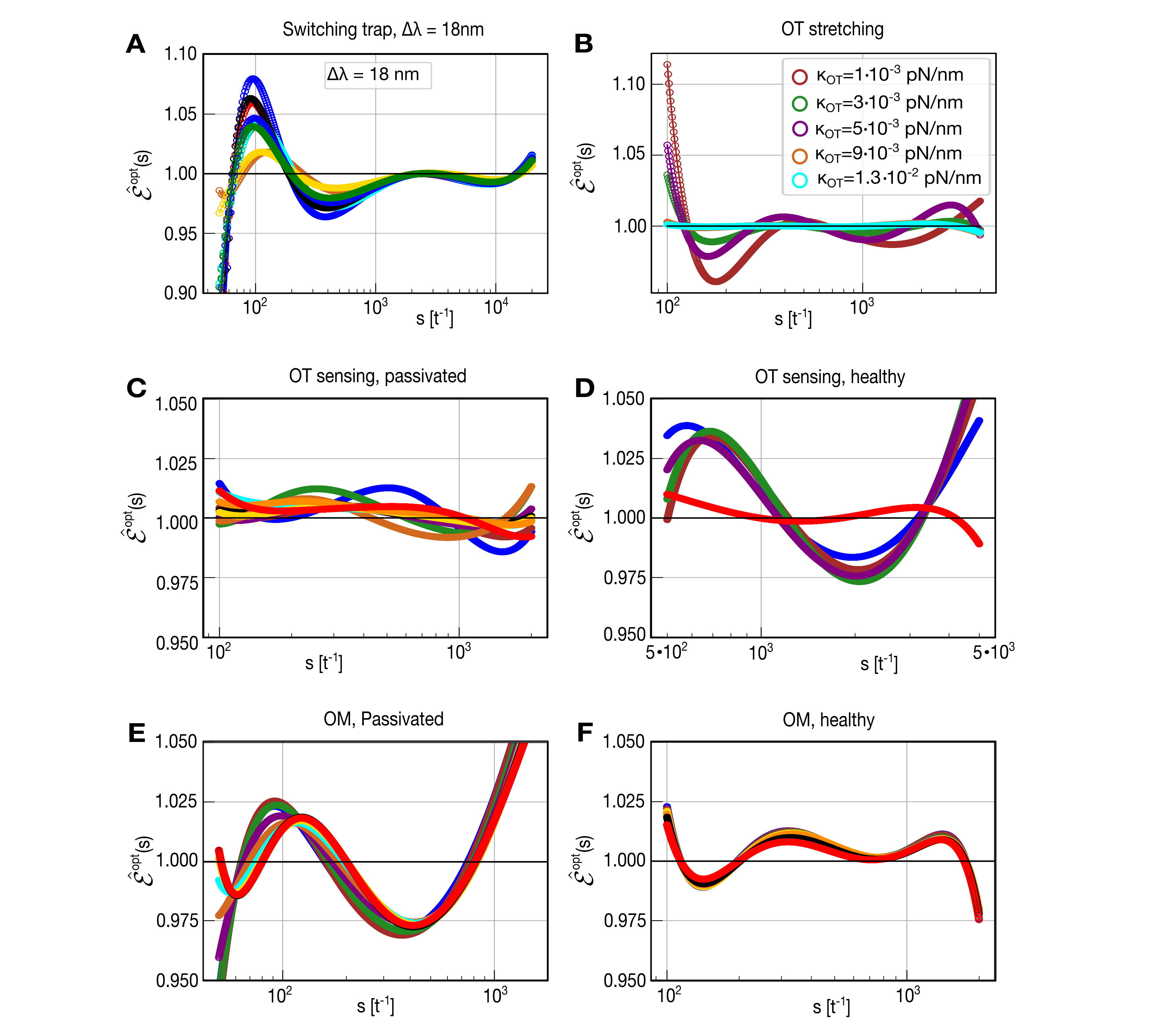} 
   \caption{\textbf{ Fits of optimal curves $\hat{\mathcal{E}}^{opt}(s)$.} Curves in the Laplace domain are obtained from the fitting procedure. Each colored curve corresponds to an experimental realization. For the SST setup (panel A), the residual $\hat{E}(s)=|\hat{\mathcal{E}}^{opt}(s)-1|$ is almost always below $0.1$, whereas for RBC experiments (panels B, C, D, E, F) it is always below $0.05$.}
   \label{fig:S2}
\end{figure}

We introduce a two-layer model with one hidden variable $y_t$ for the membrane-cortex attachment and a measurable variable $x_t$ for the membrane outer layer, Fig.~\ref{fig:S1_RBC}. Dynamics follows the equations,
\begin{align}\label{eq:ABPRBC}
    \dot{x}_t &= \mu_x\bigl(-k_b x_t - k_{\text{int}}(x_t-y_t)+C_1\bigr)+\sqrt{2D_x}\eta^x_t\\
    \dot{y}_t &= \mu_y\bigl(-k_c y_t + k_{\text{int}}(x_t-y_t)+f_t^a+C_2\bigr)+\sqrt{2 D_y }\eta^y_t\nonumber
\end{align}
where $k_b,k_c,k_{\text{int}}$ are effective stiffnesses related to the immobilizing procedure of the RBC (bare stiffness, $k_b$); the internal RBC rigidity ($k_c$); and the membrane-cortex coupling ($k_{\text{int}}$). $\mu_x,\mu_y$ and $\eta_x,\eta_y$ are bare mobilities and white noises for $x$ and $y$, while $C_1,C_2$ are just constants and $D_{x,y} =  k_B T \mu_{x,y}$. $f_t^a$ is the stochastic active force as in Section \ref{S6:RedVSR} and Eq.\eqref{eq:ABP} in the main text, obeying the following stochastic differential equation,
\begin{equation}
\label{ABP_fa}
    \dot f_t^a = -f_t^a /\tau_{a} + \sqrt{2\epsilon^2/\tau_{a}}\,\eta^{f}_t \, ,
\end{equation}
where $\overline{\eta^{f}_{t'}\eta^{f}_{t''}}=\delta(t'-t'')$. Correlation functions can be calculated using standard techniques in terms of initial conditions, which read
\begin{subequations}
\label{eq:static_correlations}
\begin{align}
 C_{xx}(0) & = \mathcal{V}_{x} =
 \frac{k_{B}T k_{y}}{k_{x} k_{y}-k_{\text{int}}^2}+\frac{\epsilon ^2 \tau_{a} k_{\text{int}}^2 \mu_{x} \mu_{y} (w^{x}_{r} \tau_{a} +w^{y}_{r} \tau_{a} +1)}{(k_{x} k_{y}-k_{\text{int}}^2)(w^{x}_{r}+w^{y}_{r}) \left((1+w^{y}_{r} \tau_{a})(1+w^{x}_{r}\tau_{a}) -k_{\text{int}}^2\, \mu_x \mu_y \tau_{a}^2\right)} \, ,\label{eq:static_correlations_a}\\[7pt]
C_{yy}(0) & = \frac{k_{B}T k_{x}}{k_{x} k_{y}-k_{\text{int}}^2}+\frac{\epsilon ^2 \tau_{a}  \mu_{y} \left(k_{x} (w^{x}_{r} \tau_{a} +1) (w^{x}_{r}+w^{y}_{r})-k_{\text{int}}^2 \mu_{y}\right)}{(k_{x} k_{y}-k_{\text{int}}^2)(w^{x}_{r}+w^{y}_{r}) \left((1+w^{y}_{r} \tau_{a})(1+w^{x}_{r}\tau_{a}) -k_{\text{int}}^2\, \mu_x \mu_y \tau_{a}^2\right)}\, , \\[7pt]
C_{ff}(0) & = \epsilon^{2} \, ,\\[7pt]
C_{xy}(0) & = \frac{k_{B}T k_{\text{int}}}{k_{x} k_{y}-k_{\text{int}}^2} +  \frac{\epsilon ^2 \tau_{a} 
k_{\text{int}}w^{x}_{r} \mu_{y} (w^{x}_{r} \tau_{a} +w^{y}_{r} \tau_{a} +1)}{(k_{x} k_{y}-k_{\text{int}}^2)(w^{x}_{r}+w^{y}_{r}) \left((1+w^{y}_{r} \tau_{a})(1+w^{x}_{r}\tau_{a}) -k_{\text{int}}^2\, \mu_x \mu_y \tau_{a}^2\right)}\, ,\\[7pt]
C_{xf}(0) & = \frac{\epsilon ^2 \tau_{a}^2 k_{\text{int}} \mu_{x} \mu_{y} }{(1+w^{y}_{r} \tau_{a})(1+w^{x}_{r}\tau_{a}) -k_{\text{int}}^2\, \mu_x \mu_y \tau_{a}^2}\, ,\\[7pt]
C_{yf}(0) & = \frac{ \epsilon ^2 \tau_{a} \mu_{y}  (w^{x}_{r} \tau_{a} +1)}{(1+w^{y}_{r} \tau_{a})(1+w^{x}_{r}\tau_{a}) -k_{\text{int}}^2\, \mu_x \mu_y \tau_{a}^2} \, 
\end{align}
\end{subequations}
and where $k_x=k_b +k_{\text{int}};k_y=k_c+k_{\text{int}}; k_{x} \mu_{x} = w^x_r; k_{y} \mu_{y} = w^y_r$. Note that constants $C_1$ and $C_2$ are not present in the connected correlation functions, as expected.
The time-dependent correlation functions can then be computed through standard techniques in Laplace space. The model \eqref{eq:ABPRBC} satisfies a reduced-VSR Eq.\eqref{eq:VSRABP_supp} (Eq.\eqref{eq:VSRABP} in the main text), which is used to fit the experimental RBC traces in Laplace space. Finally, the entropy production rate $\sigma$ can be calculated,
\begin{equation}
    \sigma_{\rm th} = \frac{\mu_y \epsilon^2(1+w^{x}_{r}\tau_{a})}{(1+w^{y}_{r} \tau_{a})(1+w^{x}_{r}\tau_{a}) -k_{\text{int}}^2\, \mu_x \mu_y \tau_{a}^2}\, ,
    \label{eq:sigmaRBC}
\end{equation}
Notice that, from \eqref{eq:static_correlations_a} and \eqref{eq:sigmaRBC} we can write $\mathcal{V}_x = \mathcal{V}_x^{\rm passive} +\alpha(\tau_{a})\sigma(\tau_{a})$ with $\mathcal{V}_x^{\rm passive} = k_{B}T k_{y}/(k_{x} k_{y}-k_{\text{int}}^2)$, $\sigma(\tau_{a})=\sigma_{\rm th}$ and 
\[ \alpha(\tau_{a}) = \frac{ \tau_{a} k_{\text{int}}^2 \mu_{x} (w^{x}_{r} \tau_{a} +w^{y}_{r} \tau_{a} +1)}{(1+w^{x}_{r}\tau_{a})(k_{x} k_{y}-k_{\text{int}}^2)(w^{x}_{r}+w^{y}_{r})} \]
We have fit the reduced-VSR to the RBCs data obtained in the optical tweezers (OT) experiments Fig.~\ref{fig:FIG4}D,E (main text) and in the ultrafast optical microscopy (OM) experiments, Fig.~\ref{fig:FIG4}F (main text). The values for $\sigma$ have been obtained from \eqref{eq:sigmaRBC} using the fit parameters obtained for each experimental trace. Fitting parameter values averaged over all experiments are shown in Tables \ref{tab:2},\ref{tab:3},\ref{tab:4}.

\section{Fitting procedure}\label{S8:Fits}
\begin{figure}[t] 
   \centering
   \includegraphics[width=\linewidth]{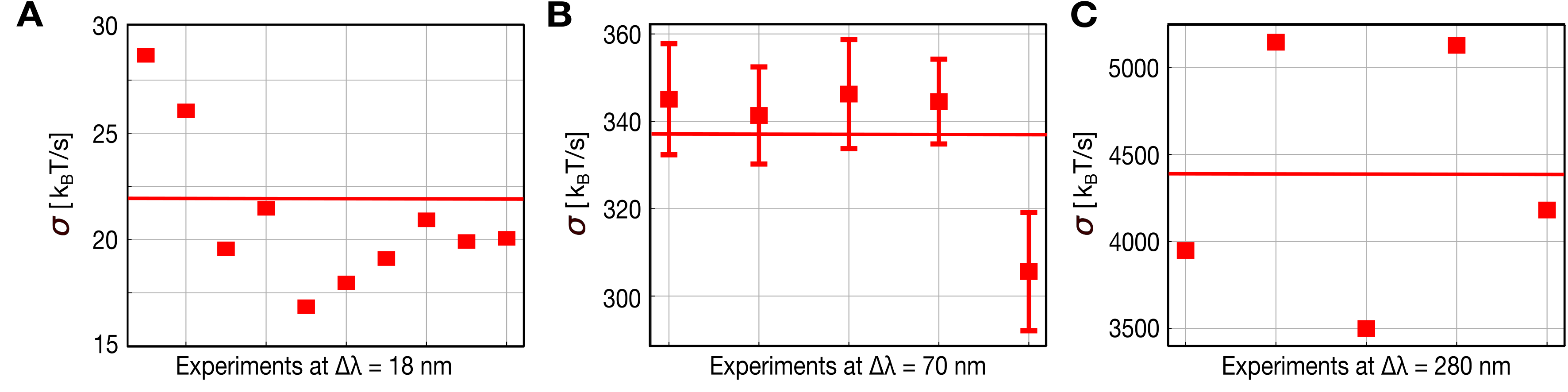} 
   \caption{\textbf{Entropy production rate for SST experiments.} Estimates of $\sigma$ for the SST setup with $\Delta\lambda= 18, 70, 280$ nm. Each point corresponds to an experimental realization, whereas the red line is the average taken over all $\sigma$ estimates for a given $\Delta\lambda$.}
   \label{fig:S3}
\end{figure}

\begin{figure}[t] 
   \centering
   \includegraphics[width=0.6\linewidth]{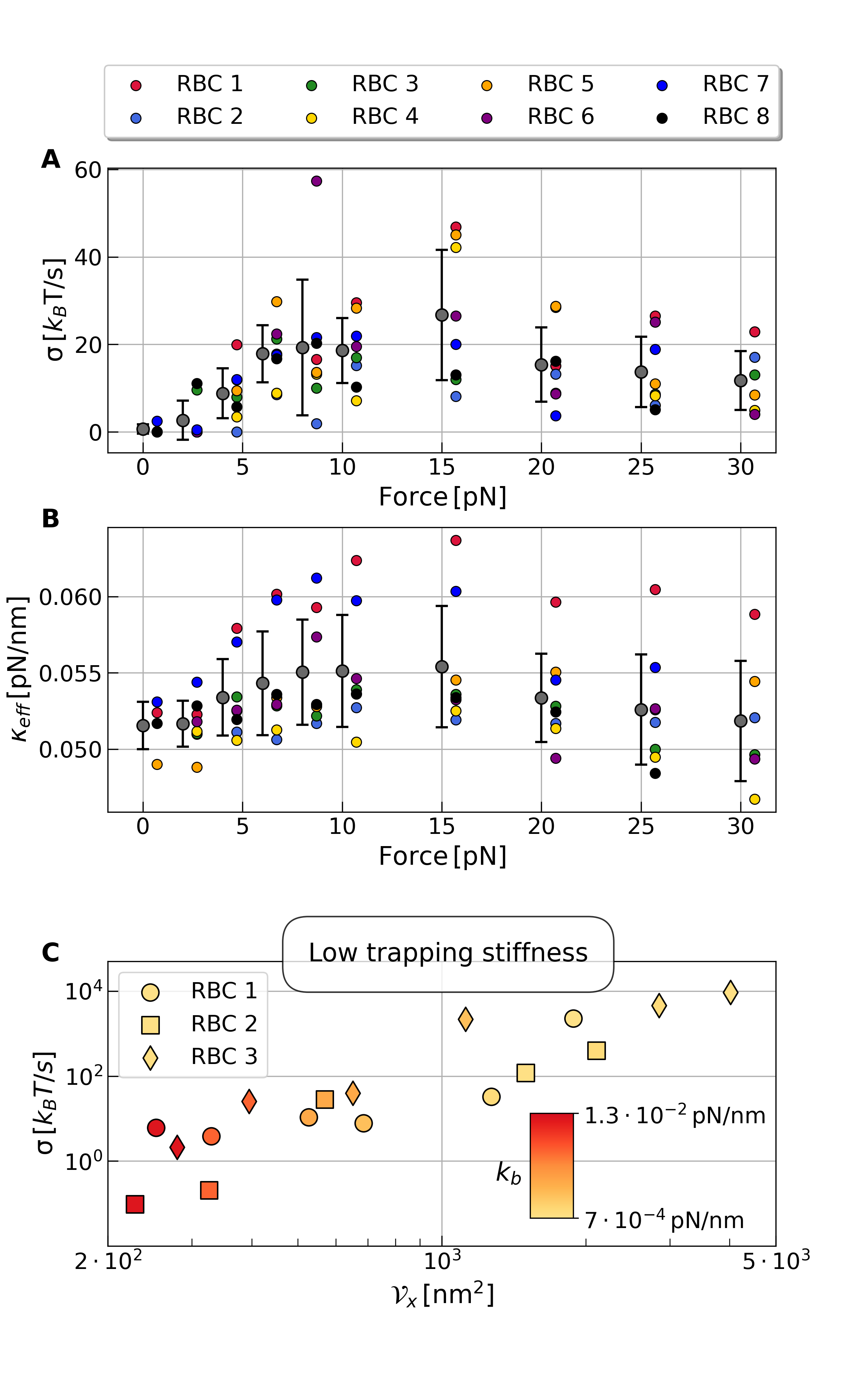} 
   \caption{\tcr{\textbf{Entropy production rate $\sigma$ for OT-stretching experiments.} (A) Measurements at high trapping stiffness ($k_{\rm b}=5\cdot 10^{-2}$pN/nm) at difference forces. $\sigma$ remains low because, at high trap stiffness, the bead's passive fluctuations mask the RBC's active fluctuations. (B) The effective trap stiffness of the two-layer active model, $k_{\rm eff}=k_x(1-\frac{k_{int}^2}{k_yk_x})$, determines the amplitude of passive fluctuations, $C_{xx}^{\rm passive}(0)=k_BT/k_{\rm eff}$ in Eq.\eqref{eq:static_correlations}. In addition, $k_{\rm eff}=k_{\rm b}+k_{\rm RBC}$ as the springs represented by the bead in the optical trap and the RBC are in a parallel configuration. The value of $k_{\rm RBC}$ was determined from the high stiffness measurements and used as a constraint for all fits at varying $k_b$ to determine the rest of two-layer model parameters $k_x,k_y,k_{\rm int},\mu_x,\mu_y,\epsilon,\tau$. (C) Color map showing the correlation between $\sigma$ and position variance ${\cal V}_x$, for low trap stiffnesses, $k_b<1.3\cdot 10^{-2}$.}}
   \label{fig:S4}
\end{figure}

We have fitted the reduced VSR Eq.\eqref{eq:VSRABP_supp} to the experimental data to determine the model parameters and derive $\sigma$. The terms in the lhs of \eqref{eq:VSRABP_supp}, ${\cal V}_{\Delta x}(t)$ and ${\cal V}_{\Sigma_x}(t)$, can be directly evaluated from the experimental traces. To perform the fit we have transformed the reduced-VSR Eq.\eqref{eq:VSRABP_supp} to the Laplace domain. Our fitting strategy is based on searching for optimal parameters that fulfill the following identity, 
\begin{equation}
   \mathcal{E}(s)=\frac{\hat{{\cal V}}_{\Delta x}(s)+\mu^2k^2\,\hat{{\cal V}}_{\Sigma_x}(s)-\hat{{\cal \tilde{S}}}(s)}{2D/s^2}=1 \, .
\label{eq:VSRABP_supp2}
\end{equation}
We denote by $\hat{\mathcal{E}}^{opt}(s)$ the optimal function obtained for the best fit parameters. Fits have been performed with the SciPy package. Fig.~\ref{fig:S2} shows different examples of $\mathcal{E}^{opt}(s)$ obtained for the SST model (Sec.\ref{S5:SST}) applied to the experimental data (panel A) and the two-layer model (Sec.\ref{S7:RBCmodel}) applied to RBC data in different experimental setups (panels B-F). Furthermore, to enhance the robustness of the fits, 
we have done a simultaneous fit of 
$\hat{{\cal V}}_{\Delta x}(s)$ and Eq.\eqref{eq:VSRABP_supp2}. \\
Optimal fit parameters are shown in Table \ref{tab:1} for the SST experiments and in Tables \ref{tab:2},\ref{tab:3},\ref{tab:4} for the RBC experiments. Parameter values are averages over all experimental realizations together with their statistical errors. Errors for $\sigma$ have been obtained by propagating the errors of the parameters entering in $\sigma_{\rm th}$ taking into account the correlations between the different parameters (see, for instance, Fig.\ref{fig:S3},\ref{fig:S4}).

\section{Results of the fits}\label{sec:fit_resuts}
Here, we present the analysis of the fits as defined in Eq.\eqref{eq:VSRABP_supp2}. Fitting the equality ${\cal E}(s)=1$ is equivalent to determining the residuals of the reduced-VSR. 

\subsection{Stochastic switching trap experiments (SST)}
\label{subsec:SSTsupp}
For the SST experiments, we have implemented a protocol in which the optical trap switches between two fixed positions separated by $\Delta \lambda$ at a rate $w$, by applying a voltage signal to a piezo actuator that controls the laser beam position. A 3 $\mu$m diameter polystyrene bead suspended in distilled water is captured in the optical trap and is used for the experiments. The bead's position in the lab frame $x_t$ is measured from the trap position $\lambda_t$ and the bead's position relative to the trap center, $y_t=F_t/k_b$, where $F_t$ is the measured force and $k_b$ the trap stiffness. Table \ref{tab:1} summarizes the parameters obtained by fitting the reduced VSR to the experimental data.

\subsection{Red blood cell (RBC) experiments}
Results for the parameters of the fits of the reduced-VSR to the optical tweezers (OT) RBC experiments are shown in Table \ref{tab:2} for OT-stretching and in  Table \ref{tab:3} for OT-sensing. Results for the parameters of the fits of the reduced-VSR to the optical microscopy (OM) experiments are shown in Table \ref{tab:4}.

\section{$\,\,$OT-sensing simulations}\label{sec:simulations}
We have considered the 3D model of Reference \cite{turlier2016equilibrium} where the OT-sensing setup was simulated to reproduce the flickering spectra of RBCs and the violation of the fluctuation-dissipation theorem. Figure \ref{fig:S11} shows a representation of the data points in which the experimental setup has been decomposed (overview, left panel): the RBC (red points), the probe bead (dark blue points) and the three additional beads used to stabilize the RBC position (light blue points). The middle and right panels show a zoom of the overview showing the points of the RBC in contact with the probe bead for active (middle panel) and passive (right panel). The blue points are taken from t=0 in a time series of 10000 frames (active) and 15000 frames (passive) with sampling rate every 1.33$\cdot 10^{-4}$s for a total time of 1.33s (active) and 2s (passive). The yellow-orange points near the center of the probe bead indicate the positions of the geometric center of the bead. The red points are also taken from t=0 in a time series of 200 frames (active) and 300 frames (passive) with sampling rate every 6.6$\cdot 10^{-3}$s for a total time of 1.33s (active) and 2s (passive).  We have applied the reduced-VSR and estimated $\sigma$ from the trajectory of the probe bead (yellow-orange points) projected on the x-axis along the normal of the RBC membrane, mimicking the experimental conditions. Figure \ref{fig:S12} shows the position trajectory $x_t$ and its distribution $P(x_t)$ in passive (top panels) and active (bottom panels). As can be seen, the variance is larger for the active than the passive. Although the time step is longer for simulations than for experiments (1.33$\cdot 10^{-4}$s versus 4$\cdot 10^{-5}$s), the reduced-VSR rule fits the data well, giving $\sigma\sim 10000 k_BT/s$ for the active RBC and $\sigma\sim 25 k_BT/s$  for the passive RBC, values not far from the OT-sensing data (Table \ref{tab:5}, Figure \ref{fig:FIG4}E). 

\section{$\,\,$Structure factor for $\sigma$-map.}\label{sec:structure_factor}
\begin{figure}
   \centering
\includegraphics[width=1\linewidth]{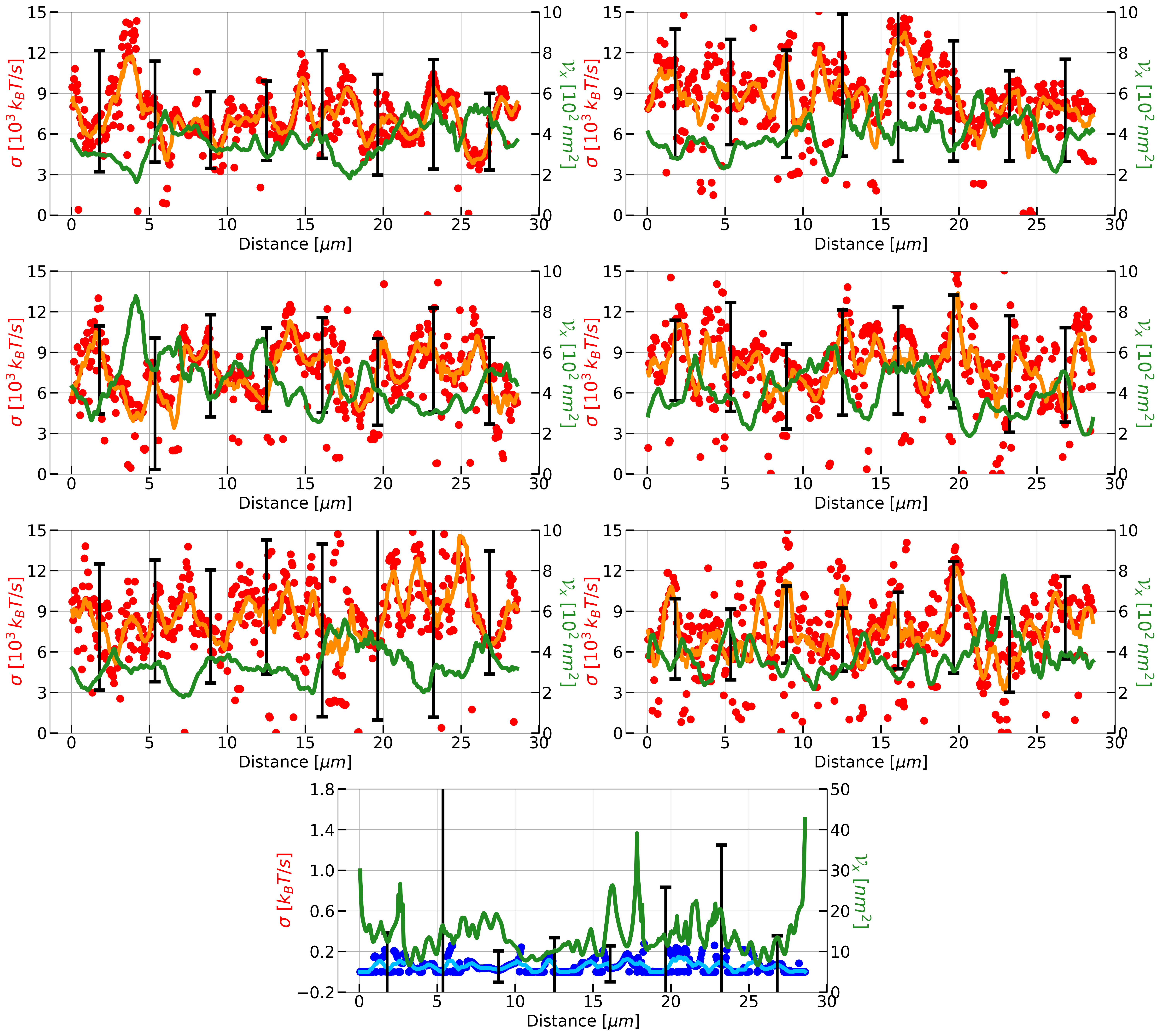} 
   \caption{\textbf{Profiles of the entropy production rate $\sigma$ and the variance of fluctuations ${\cal V}_x$ for OM experiments with RBCs.} $\sigma$ in $\text{k}_{B}\text{T}/\text{s}$ (red points, left scale) and variance of fluctuations (green line, right scale in $nm^2$) for six healthy RBCs. The orange line is a running average. The lower panel is a passivated RBC, blue points (left scale) are $\sigma$ values, and green (right scale) is the variance. The cyan line is a running average over $\sigma$. Statistical errors are shown as black lines.}
   \label{fig:RBC1-5}
\end{figure}

\begin{figure}
   \centering
   \includegraphics[width=\linewidth]{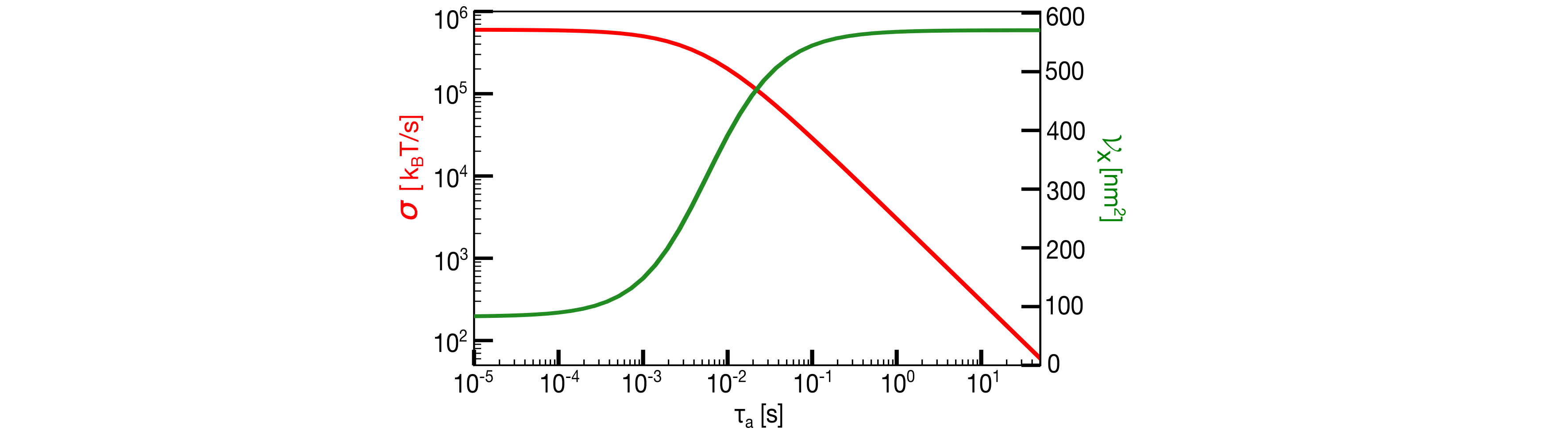} 
   \caption{\textbf{Analytical dependence of $\sigma$ and ${\cal V}_x$ on the correlation time $\tau_{a}$ of the active noise $f^a_t$.} The entropy production rate $\sigma$ (red curve, left axis) decreases with $\tau_{a}$, while the variance of the position $\mathcal{V}_{x} = \overline{x^2}-\overline{x}^2$ (green curve, right axis) increases with $\tau_a$.
   Other parameters are fixed to $k_x = 5\cdot10^{-2}$ pN/nm, $\mu_x=2\cdot 10^4$ nm/(pN s), $k_y = 1\cdot10^{-2}$ pN/nm, $\mu_y=2\cdot 10^4$ nm/(pN s), $k_{\text{int}} = 1\cdot10^{-2}$ pN/nm, $\epsilon^2=30$ $\text{pN}^2$, similar to experimental values in Table \ref{tab:4}. The behavior depicted here causes the anti-correlation between $\sigma$ and ${\cal V}_x$ discussed in the main text and shown in Fig.~\ref{fig:FIG4}G.}
   \label{fig:S6}
\end{figure}

\begin{figure}
   \centering
   \includegraphics[width=0.7\linewidth]{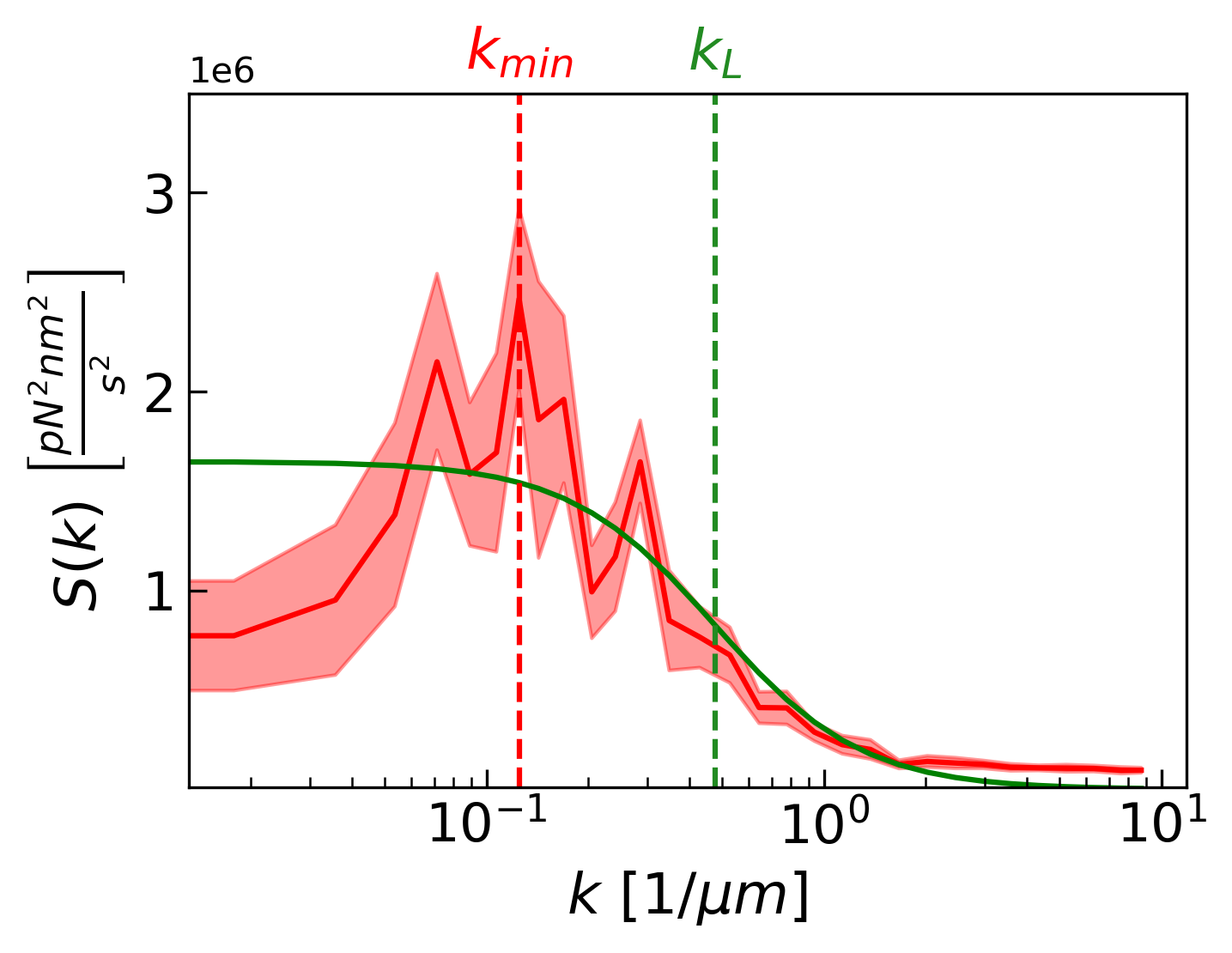} 
   \caption{\textbf{Heterogeneity analysis.} Structure factor (red line) of the $\sigma$-field along the cell contour averaged over six RBC. The red area around the red line represents the standard error. The smooth green line is the Lorentzian fit.}
   \label{fig:S7}
\end{figure}

Fig.~\ref{fig:FIG4}F shows that $\sigma$ is very heterogeneous along the RBC contour. A possible explanation is the underlying heterogeneous cortex-membrane binding-unbinding dynamics that produces domains of different activity $\sigma$. Such heterogeneity can be quantified by measuring the structure factor $S(k)$ of the $\sigma$ field along the cell contour, i.e., the Fourier transform in reciprocal space of the spatial correlation function $C_{\sigma\sigma}$ shown in Fig.~\ref{fig:FIG4}H.
We have $S(k)=\langle |\hat{\sigma}_{k}|^2 \rangle= \sum_{p=0}^{N-1}e^{2\pi i p k /N}\sigma_{p}$, $k=0,1,...,N-1$, where $\sigma_p$ is the $\sigma$ value at pixel $p$ and $N=512$ is the total number of pixels along the cell contour. Fig.~\ref{fig:S7} shows the results averaged over six RBCs. As we can see, there is a peak at a minimum $k_{{\rm min}}$ value, observed for each RBC, which gives a characteristic domain length $l=(2\pi k_{{\rm min}})^{-1}\approx 1.3 \mu m$. The value of $l$ is larger than the characteristic correlation length of the $\sigma$ field, $\xi_{\sigma \sigma}\approx 0.35 \mu m$. This result shows that the correlation length of the active $\sigma$ field $\xi_{\sigma \sigma}$ is lower than the typical size $l$ of the heterogeneous $\sigma$-domains. Notice that $\xi_{\sigma \sigma}$ equals the inverse of the characteristic wavevector $k_{{\rm L}}$ of the Lorenzian fit (Fig.~\ref{fig:S7}, smooth green line) showing two distinct characteristic lengths for the $\sigma$-field. The size $l$ of the $\sigma$-domains correlates with the ruggedness scale, observed in Figure R1 and the panels of Fig.~\ref{fig:RBC1-5}, which falls in the micrometer range. Albeit less reproducible over different RBCs, another peak in $S(k)$ is observed in Figure \ref{fig:S7} for lower values of $k\approx 0.04\mu^{-1}$  corresponding to a larger length scale of about $4\mu m$, showing that the $\sigma$-map contains useful information about the dissipation length scales.

\section{$\,\,$Active two-layer model in a ladder}\label{sec:ladder}

\begin{figure}
   \centering
   \includegraphics[width=0.45\linewidth]{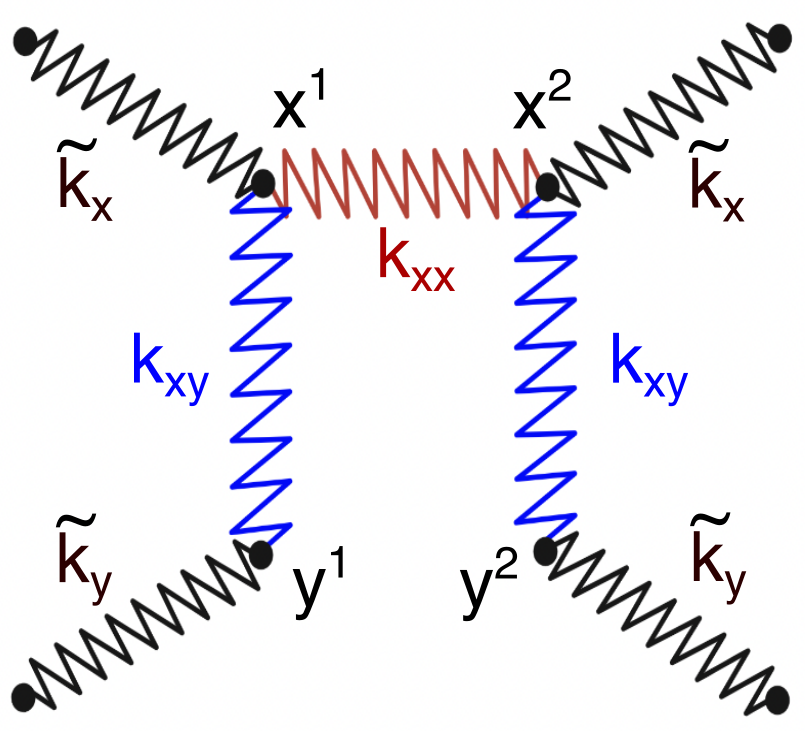} 
   \caption{\textbf{Schematics of the ladder two-layer model.}  The effective coupling parameters $k_x,k_y$ appearing in Eq.(S44) are combinations of the coupling parameters $\tilde k_x,\tilde k_y, k_{xy}, k_{xx}$ depicted in the figure.}
   \label{fig:S8}
\end{figure}

\begin{figure}
   \centering
   \includegraphics[width=7.1cm]{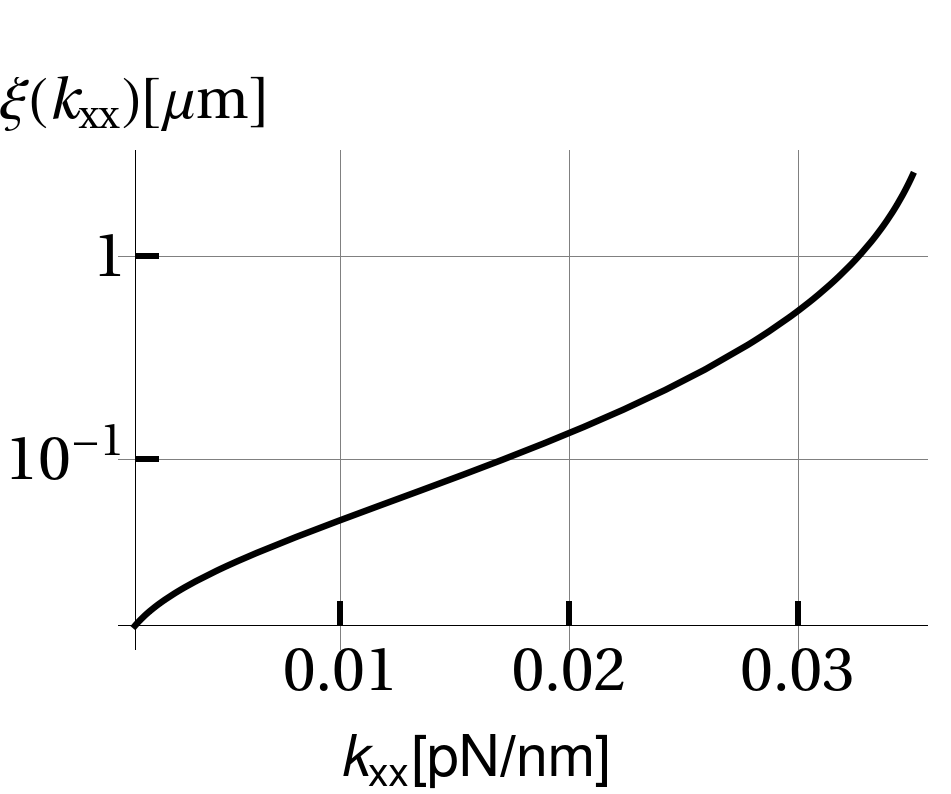}
   \hspace{0.5cm}
   \includegraphics[width=7.5cm]{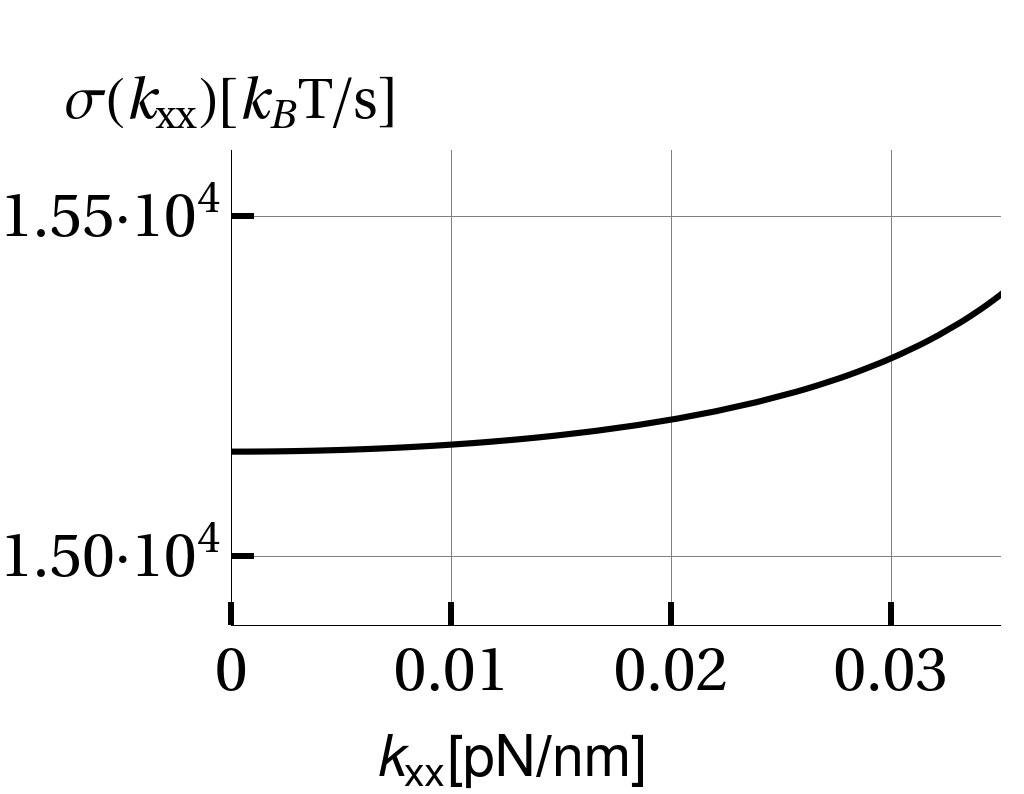} 
   \caption{\textbf{Correlation length and $\sigma$ for the ladder model.} Dependence of the correlation length $\xi(k_{xx})$ (left) and $\sigma(k_{xx} )$ (right) for the ladder two-layer model.}
   \label{fig:S9}
   
\end{figure}

To better understand the role of spatial correlations on $\sigma$, we have considered a zero-dimension ladder system, a double two-layer model described by the probes positions $x^1_t$, $x^2_t$ and hidden degrees of freedom $y^1_t$, $y^2_t$ and identical couplings $k_x$ and $k_y$ (Fig.~\ref{fig:S8}). The ladder has a lateral coupling constant $k_{xx}$, equivalent to an effective membrane tension that introduces correlations between probes at positions $x^1_t$, $x^2_t$. The following equations describe the system:
\begin{equation}
    \begin{split}
        \dot{x}^1_t &= \mu_x\bigl(-k_x x^1_t + k_{xy}y^1_t+k_{xx}x^2_t\bigr)+\sqrt{2D_x}\eta^{x^1}_t \\
        \dot{y}^1_t &= \mu_y\bigl(-k_y y^1_t + k_{xy}x^1_t+f^1_t\bigr)+\sqrt{2 D_y }\eta^{y^1}_t \\ 
        \dot{x}^2_t &= \mu_x\bigl(-k_x x^2_t + k_{xy}y^2_t+k_{xx}x^1_t\bigr)+\sqrt{2D_x}\eta^{x^2}_t \\
        \dot{y}^2_t &= \mu_y\bigl(-k_y y^2_t + k_{xy}x^2_t+f^2_t\bigr)+\sqrt{2 D_y }\eta^{y^2}_t \\
        \dot f_t^1 &= -f_t^1 /\tau_{a} + \sqrt{2\epsilon^2/\tau_{a}}\,\eta^{f^1}_t\\
        \dot f_t^2 &= -f_t^2 /\tau_{a} + \sqrt{2\epsilon^2/\tau_{a}}\,\eta^{f^2}_t \, .
    \end{split}
\end{equation}
The ladder two-layer model is exactly solvable and correlation function at equal times, $C_{x^1 x^2 } (0)$ and $C_{x^1 x^1 }(0)= C_{x^2 x^2 }(0)$, give an estimate of the correlation length for the position fluctuations, $\xi(k_{xx})=d/\log(C_{x^1x^1}(0)/C_{x^1x^2}(0))$. Here d=50nm is the distance between contiguous regions (pixels), and coupling parameters are taken from Table \ref{tab:4} for the OM measurements. The results for $\xi(k_{xx} )$ and $\sigma(k_{xx} )$ for the ladder are shown in Fig.~\ref{fig:S9}. Interestingly, we find that $\xi$ quickly grows with $k_{xx}$ falling in the micrometer range for $k_{xx}\approx0.04pN/\mu m$, which is the same range as the other coupling parameters $k_x$ and $k_y$. In contrast, $\sigma$, which can be calculated and is equal to
\begin{equation}
\begin{split}
    \sigma =& \frac{2 \epsilon^2 \mu_y \mu_x \tau_a}{\left(\tau_a \left(k_x (k_y \mu_x \mu_y \tau_a+\mu_x)-\mu_x \left(\mu_y \tau_a \left(k_{xx} k_y+k_{xy}^2\right)+k_{xx}\right)+k_y \mu_y\right)+1\right)}\cdot\\[7pt]
    &\hspace{0.5cm}\cdot\bigg(\frac{  k_x^2 \mu_x \tau_a (k_y \mu_y \tau_a+1)+k_x \left(\mu_y \tau_a \left(2 k_y-k_{xy}^2 \mu_x \tau_a\right)+2\right)}{\left(\mu_x \tau_a \left(k_y \mu_y \tau_a (k_x+k_{xx})+k_x+k_{xx}-k_{xy}^2 \mu_y \tau_a\right)+k_y \mu_y \tau_a+1\right) }\\[7pt]
    &\hspace{0.5cm}-\frac{\tau_a \left(k_{xx}^2 (k_y \mu_x \mu_y \tau_a+\mu_x)+k_{xy}^2 \mu_y\right)+(1+k_y \mu_y \tau_a)/\mu_x \tau_a}{\left(\mu_x \tau_a \left(k_y \mu_y \tau_a (k_x+k_{xx})+k_x+k_{xx}-k_{xy}^2 \mu_y \tau_a\right)+k_y \mu_y \tau_a+1\right) } \bigg)
    \end{split}
\end{equation}
remains nearly independent of $k_{xx}$ and $\xi$, demonstrating that the measured correlation length $\xi_{xx}\approx 0.8\mu m$ in Fig.~\ref{fig:FIG4}H (green line) is very sensitive to small $\sigma$ variations. This result supports the idea that active systems are critical (59,60), and a small change in $\sigma$ leads to a large increase in the correlation length $\xi_{xx}$. Albeit simplistic, the zero-dimensional ladder model already shows the importance of building spatial interactions in refined descriptions of the $\sigma$ field, possibly by including disorder in the model.

\section{$\,\,$VSR-based fits vs. power spectrum-based fits}\label{sec:VSRvsPS}
\begin{figure}
   \centering
   \includegraphics[width=0.8\linewidth]{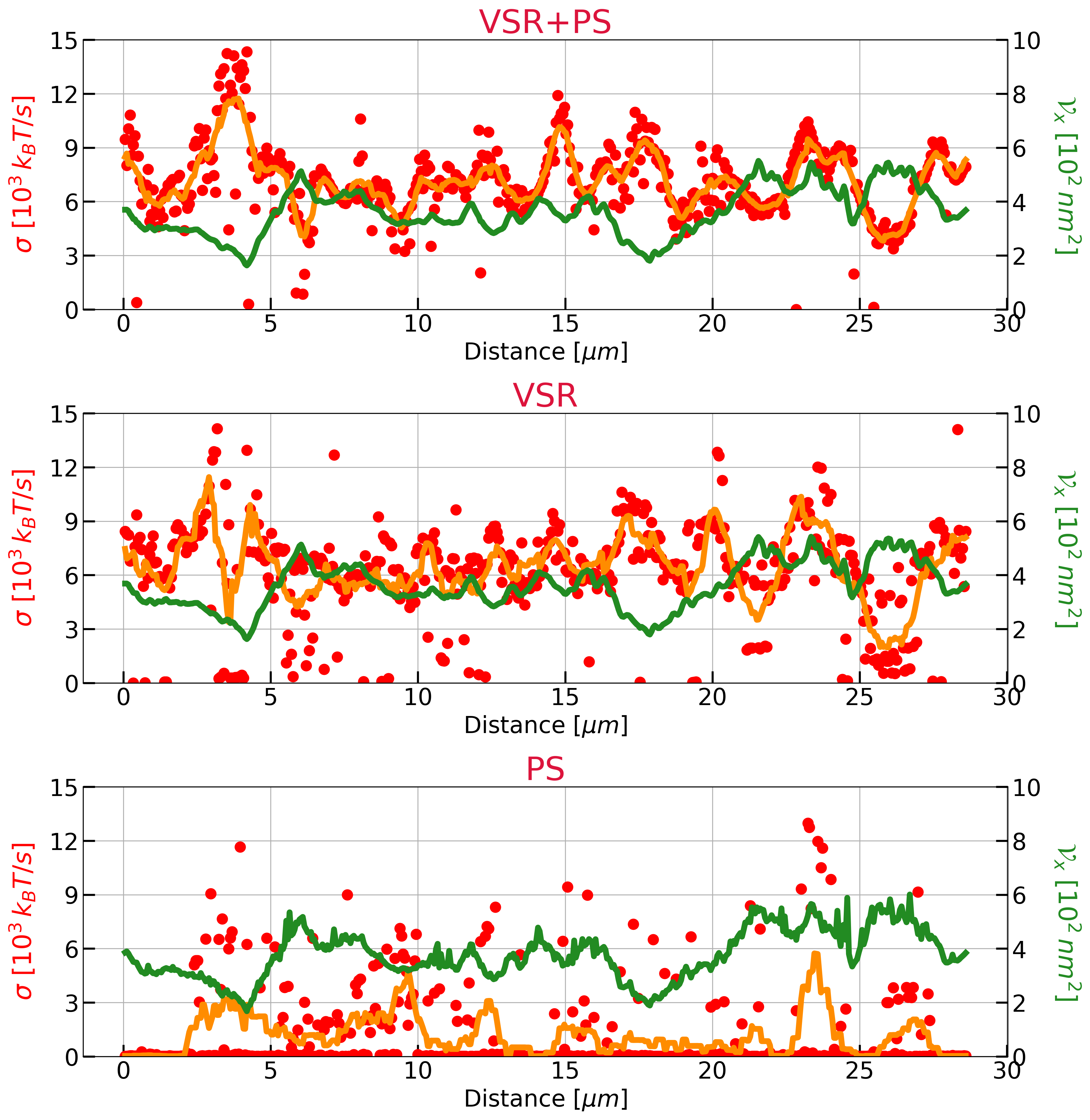} 
   \caption{\textbf{Comparison between reduced-VSR  fitting and variance of the displacement fitting.} $\sigma$-maps for the OM experiments produced by fitting the VSR and the power spectrum, like in the main text (A), the VSR alone (B), and the power spectrum alone (C).}
   \label{fig:S10}
\end{figure}

\begin{figure}
   \centering
\includegraphics[width=\linewidth]{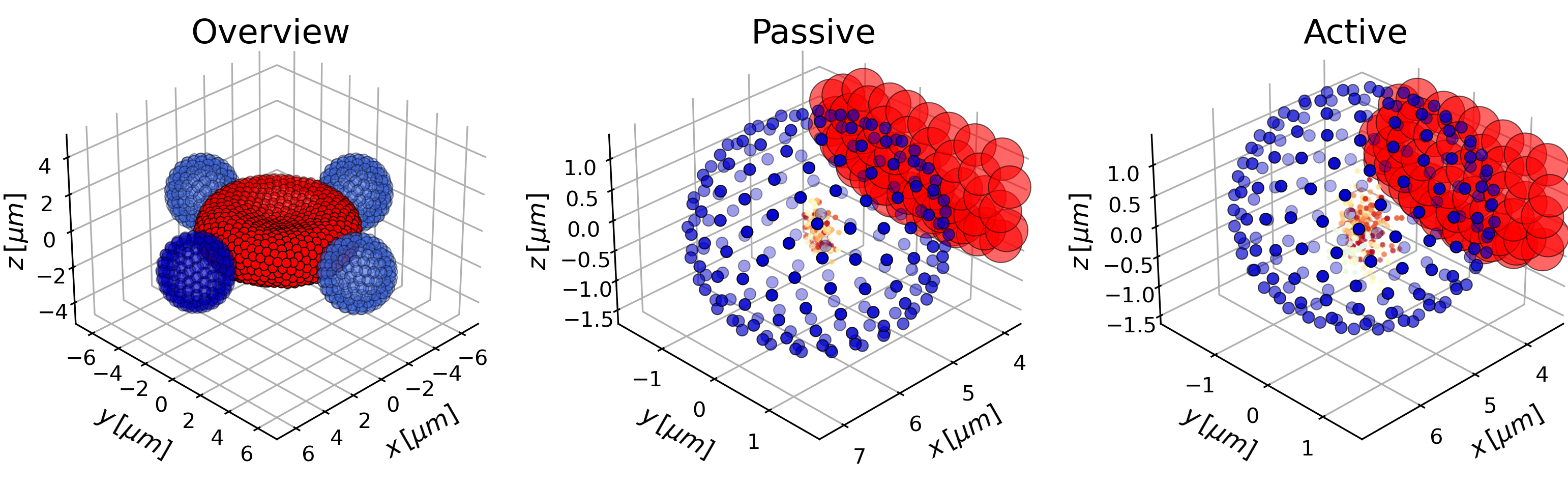} 
    \caption{\textbf{Visualisation of simulated data.} Schematic of the 3D simulation model from Ref. \cite{turlier2016equilibrium} for the OT-sensing experiments. (Left panel) The RBC (red) is attached to four beads (blue), one bead is the measurement probe (dark blue), the other three beads stabilize the RBC (light blue). (Middle and right panels) Zoomed region between the probe bead and the RBC for passive and active case. The colored points indicate the bead's geometric center.}
   \label{fig:S11}
\end{figure}

\begin{figure}
   \centering
\includegraphics[width=0.8\linewidth]{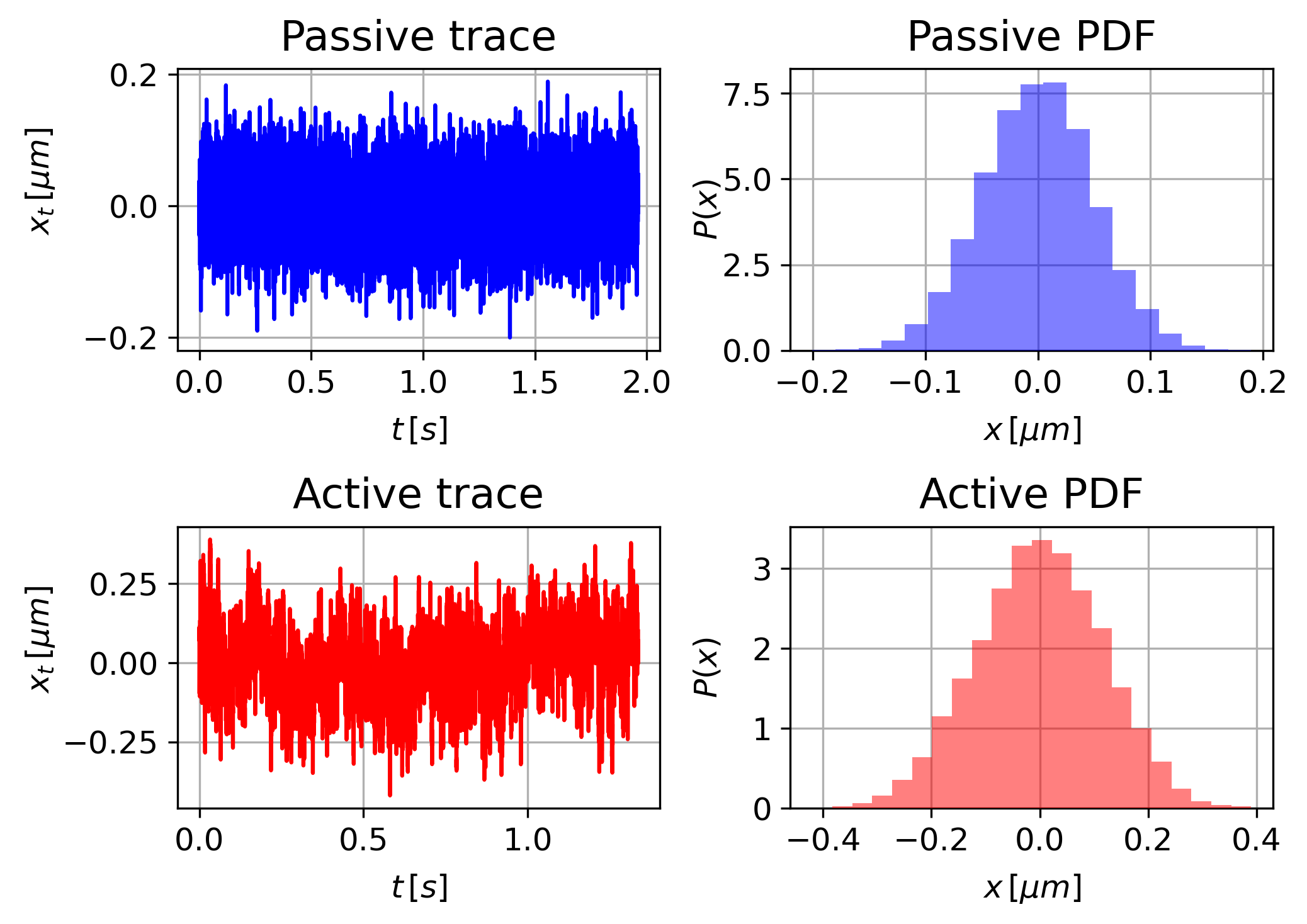} 
   \caption{\textbf{Traces and PDF for simulated data.} Simulation trajectories and position distribution for the $x_t$ coordinate of the bead's geometric center in the active and passive RBC for the simulation of the OT-sensing experiment taken from Ref.\cite{turlier2016equilibrium}}
   \label{fig:S12}
\end{figure}

As pointed out in Section \ref{S8:Fits}, we have also fit the Laplace transform of the displacement variance $\hat{\mathcal{V}}_{\Delta x}(s)=2(\hat{C}_{xx}(s)-C_{xx}(0)/s)$ to the experimental data to get more robust fits. $\hat{\mathcal{V}}_{\Delta x}(s)$ contains the same amount of information as the power spectral density (PSD) $\hat{C}_{xx}(\omega)$, but in the Laplace domain $s$, which is the quantity usually fitted to experimental data,  as done for example in \cite{tucci2022modeling}. We point out that the fits are performed in Laplace space because linear models are easily solved in this domain, and correlation functions have simple polynomial behavior. In contrast, they can be very complicated in real space. In this section, we aim to compare the performances of fits based on the reduced-VSR \eqref{eq:VSRABP} and fits based on the variance of the displacement $\hat{\mathcal{V}}_{\Delta x}(s)$, equivalent to fit the PSD $\hat{C}_{xx}(\omega)$. In Fig.~\ref{fig:S10}, we show the results of this comparison: Fig.~\ref{fig:S10}B shows fits using the reduced-VSR only; Fig.~\ref{fig:S10}C shows fits only using the PSD in Laplace space, i.e., $\hat{\mathcal{V}}_{\Delta x}(s)$. Both fits start from the same initial seeds, fit parameters bounds, and fit ranges as in Fig.~\ref{fig:S2}E-F. We immediately see that the reduced-VSR fits in Fig.~\ref{fig:S10}B are compatible with the results of the simultaneous fits to the reduced-VSR and $\hat{\mathcal{V}}_{\Delta x}(s)$ shown in the main text (Fig.~\ref{fig:S10}A). Moreover, the reduced-VSR fits give an accurate $\sigma$-map (Fig.~\ref{fig:S10}B), which is nonexistent in Fig.~\ref{fig:S10}C.

\section{$\,\,$Nonlinear models}\label{sec:nonlinear}
\tcr{
As a further example of a system with non-Gaussian statistics, we consider the RBC model \eqref{eq:ABPRBC} with $f_t^a$ obeying an autonomous Poisson process switching stochastically between two values $\pm \epsilon$ with rate $1/(2 \tau_a)$. The autocorrelation $\langle f_t^a f_0^a \rangle =  \epsilon^2 \exp(-t/\tau_a)$ is the same of the Ornstein-Ulhenbeck process \eqref{ABP_fa}. Notice that this model differs from the stochastic switching trap of Figure \ref{fig:FIG2} (main text) and Sec.\ref{subsec:SSTsupp} by the presence of an extra layer (y), which results in the dynamics of the x variable non-Gaussian and is effectively non-linear. In this regard, the model is not so different from the stochastic switching trap shown in Figure \ref{fig:FIG2}, but extends the non-linearity to an additional intermediate degree of freedom (y). For the two-layer dichotomous model, we can use the formulas derived in Section \ref{S7:RBCmodel} for analyzing data and computing the entropy production rate $\sigma$ from the estimated parameters. Figure~\ref{fig:S14} shows that the statistics of the variables are non-Gaussian. Yet, the entropy production is well estimated with the protocol described in Section~\ref{S8:Fits}.} 

\tcr{The other case we have considered is a non-linear model defined by a particle in a quartic potential aiming to capture higher-order fluctuation modes, $U(x)=\frac{1}{2}kx^2+\frac{1}{4}gx^4$, under a Gaussian active noise of amplitude $\epsilon$ and correlation time $\tau$. Here, non-linearity enters at the level of the potential through the non-linear parameter $g$. Results are shown in Figure \ref{fig:S15} for the quartic potential (red). As shown, $x_t$ traces give non-Gaussian distributions in the quartic model (red distribution in the third small panel, $P(x)$ versus $x$). In contrast, for a purely harmonic model ($g=0$) the distribution is Gaussian (blue distribution in the third small panel, $P(x)$ versus $x$). The figure also shows the distributions (4th-6th small panels at the bottom, red) of the parameters $\epsilon,\tau,g$ obtained by fitting the reduced-VSR \eqref{eq:VSRABP} and \eqref{eq:redS_comp}, with $F_I$ equal to the sum of term $-gx^3$ plus the active noise, to position traces using simulation-based inference (SBI). The method yields parameter distributions that agree with the actual values (black vertical bars). Finally, the right panel shows the estimated $\sigma$ values derived from the fitting parameters, plotted versus the "theoretical" one, obtained by applying Sekimoto's formula Eq.(S12) to simulated data.  We conclude that the VSR approach is generally valid for linear and non-linear systems.}

\clearpage

\clearpage


\clearpage

\begin{figure}
   \centering
\includegraphics[width=\linewidth]{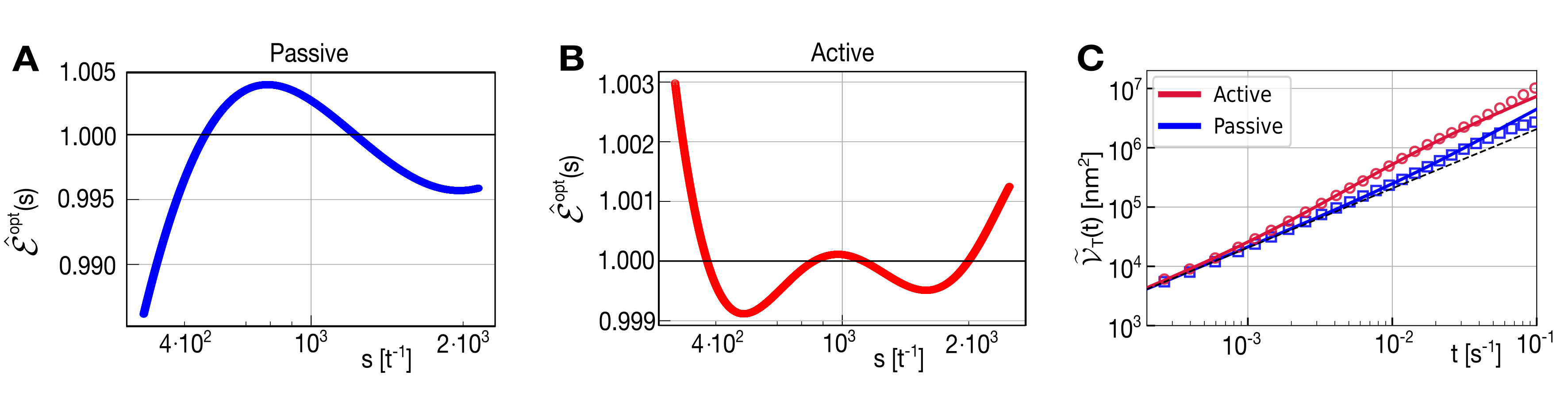} 
   \caption{\textbf{Fitting results for the OT-sensing simulations from Ref.\cite{turlier2016equilibrium}.}  We show the fitting function Eq. \eqref{eq:VSRABP_supp2}. Values of  $\sigma\sim 25k_{B}T/s$ (passive) and $\sigma\sim 10^4k_{B}T/s$ (active) are obtained (Table \ref{tab:5}). \tcr{Panel C shows the fits to $\tilde{\cal V}_T(t)={\cal V}_{\Delta x}(t)+\mu^2k^2\,{\cal V}_{\Sigma_x}(t)$ for active and passive simulations.}}
   \label{fig:S13}
\end{figure}

\begin{figure}
   \centering
\includegraphics[width=0.99\linewidth]{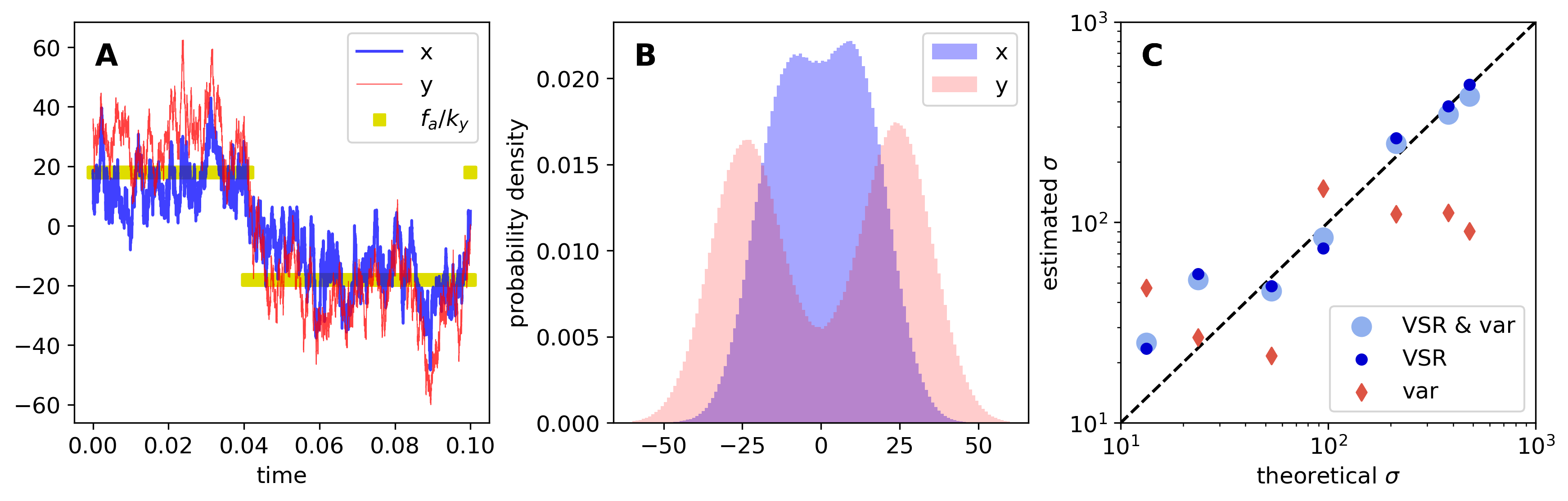} 
\caption{\tcr{\textbf{Simulations of the two-layer stochastic switching trap.} The model corresponds to Eqs.\eqref{eq:ABPRBC} with the dichotomic noise of Eq.\eqref{LE_switch_present} centered around 0 replacing the active force $f_t^a$. In dimensionless units, we  used $k_x=0.06$, $k_y=0.05$, $k_{int}=0.03$, $\mu_x=\mu_y=2\times 10^4$, $\tau_a=0.01$. Each simulation, for a given $\epsilon$ in the range $[0.15,0.9]$, collects $10^7$ samples spaced with time step $\Delta t=10^{-5}$ (while the integration time step is $5\times 10^{-8}\ll \Delta t$). Then, only the trajectory of $x_t$ is used for estimating the entropy production rate with the protocol described in Section~\ref{S8:Fits}.
   \textbf{A.} A piece of the trajectory of the three degrees of freedom for $\epsilon=0.9$, showing the jumping dynamics of the active force $f_t^a$.
   \textbf{B.} Non-Gaussian stationary probability density of the observable variable $x_t$ and of the hidden variable $y_t$. 
   \textbf{C.} 
   Entropy production rate estimated by plugging the fitted parameters into \eqref{eq:sigmaRBC} as a function of the $\sigma$ computed with the actual parameters, for three fitting procedures (see Section~\ref{S8:Fits}): fitting both the VSR and the variance of the displacement (sky blue circles), only the former (blue circles), or only the latter (red diamonds). One can note that results are good only if the VSR is included in the fit.
   }}
   \label{fig:S14}
\end{figure}

\begin{figure}
   \centering
\includegraphics[width=0.99\linewidth]{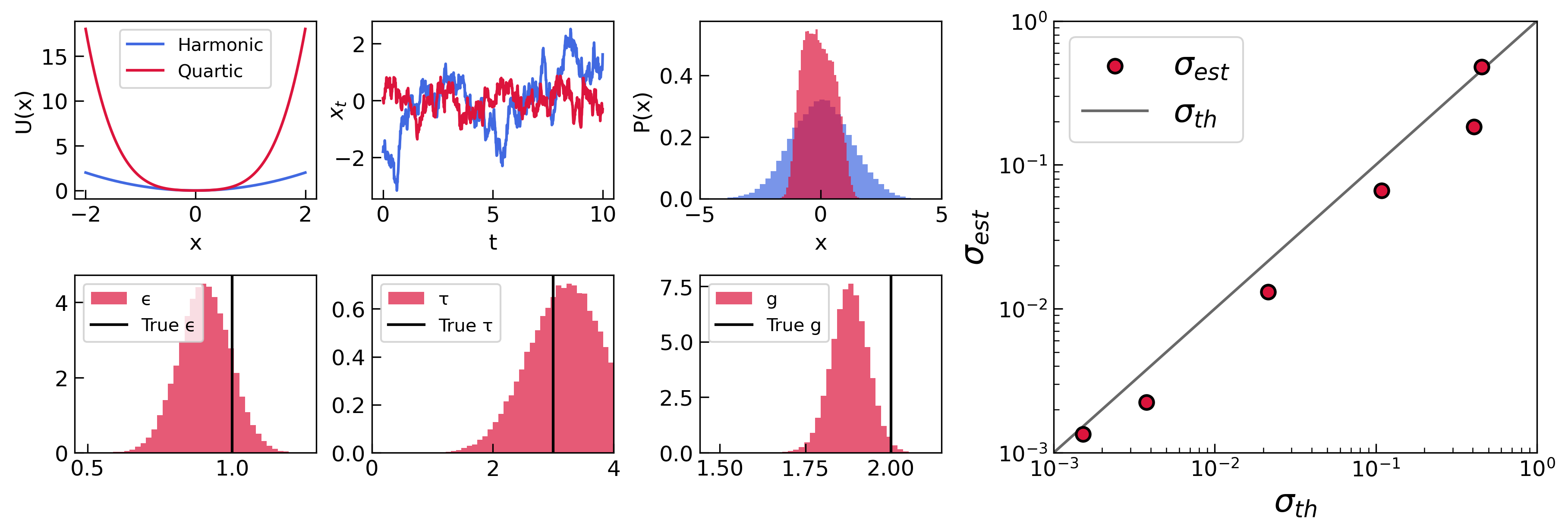} 
\caption{\tcr{\textbf {Numerical simulations of a particle in a non-linear quartic potential under Gaussian active noise.} Estimated $\sigma$ values obtained from the reduced-VSR agree with the expected values over three decades of $\sigma$.  }}
   \label{fig:S15}
\end{figure}
\clearpage

\begin{table}[bth]
    \centering
    \begin{tabular}{l|ccc}
        $\Delta\lambda$ & 18 nm& 70 nm& 280 nm\\
        \hline\\[-2ex]
        $1/w$ [$10^{-2}$s]& $5$ & $5$ & $5$ \\
        $\tau_a$ [$10^{-2}$s] & $5.0\pm 0.5$  & $5.0\pm 0.1 $  & $5.0\pm 0.5$ \\[1.5pt]
        \hline\\[-2ex]
        $\epsilon_{SST}$ [pN] & $0.5$ & $1.96$  
        & $7.84$ \\
        $\epsilon_{fit}$ [pN] & $0.5\pm 0.05$ & $1.99\pm 0.03$ & $7.8\pm 0.1$ \\[1.5pt]
        \hline\\[-2ex]
        $\mu $ [$10^{4}$ nm/(s pN)] & $4.0\pm 0.3$  & $4.0\pm 0.3$  & $4.0\pm 0.3$ \\
        $\mu_{fit} $ [$10^{4}$ nm/(s pN)] & $4.20\pm 0.01$  & $4.000\pm 0.005$ & $3.6\pm 0.2$ \\[1.5pt]
        \hline\\[-2ex]
        $k_b $ [$10^{-2}$ pN/nm] & $5.6\pm 0.3$  & $5.6\pm 0.3$  & $5.6\pm 0.3$ \\
        $k_{fit} $ [$10^{-2}$ pN/nm] & $5.45\pm 0.01$ & $5.55\pm 0.01$  & $5.55\pm 0.01$ \\[1.5pt]
        \hline\\[-2ex]
        $\sigma $ [$\text{k}_{B}\text{T/s}$] & $22\pm 4$ & $330\pm 30$ & $5300\pm 500$ \\
        $\sigma_{fit} $ [$\text{k}_{B}\text{T/s}$] & $22\pm 3 $  & $337\pm 8 $  & $4400\pm 500$ \\[1.5pt]
        \hline
    \end{tabular}
    \caption{\textbf{Fit parameters for SST experiments}. Results of the fits shown in Fig.~\ref{fig:FIG3}A (main text) with known fixed parameters $\mu=4\cdot 10^4$ nm(pN s)$^{-1}$, $k=5.6\cdot 10^{-2}$ pN/nm and $1/w=0.05 s$. $\Delta\lambda$ is taken equal to 18, 70, 280 nm and determines the strength of the active noise $\epsilon$ when the SST model is mapped into the ABP (Eq. \ref{eq:ABP}, main text). For all setups, the values of the fit parameters (lower rows) and the estimates for $\sigma$ are compatible with nominal values (upper rows) within the statistical error.}
    \label{tab:1}
\end{table}

\clearpage

\begin{table}[ht!]
    \centering
\tcr{\begin{tabular}{l|ccccc}
Force & 0 pN & 2 pN & 4 pN & 6 pN & 8 pN  \\[1.5pt]
\hline\\[-2ex]
$k_x$ [$10^{-2}$ pN/nm]& $6.0\pm 0.1$ & $6.3\pm 0.2$ & $6.9\pm 0.2$ & $7.3\pm 0.3$ & $7.8\pm 0.4$ \\[1.5pt]
\hline\\[-2ex]
$\mu_x$ [$10^4$ nm/(pN s)]& $2.1\pm 0.1$ & $2.3\pm 0.1$ & $2.3\pm 0.1$ & $2.3\pm 0.1$ & $2.3\pm 0.1$ \\[1.5pt]
\hline\\[-2ex]
$k_y$ [$10^{-1}$ pN/nm] & $1.1\pm 0.3$ & $0.9\pm 0.1$ & $1.1\pm 0.2$ & $1.6\pm 0.2$ & $1.3\pm 0.2$ \\[1.5pt]
\hline\\[-2ex]
$\mu_y$ [$10^3$ nm/(pN s)] & $2.6\pm 0.6$ & $4.3\pm 0.5$ & $5.3\pm 0.8$ & $5.3\pm 0.6$ & $8\pm 1$ \\[1.5pt]
\hline\\[-2ex]
$k_{\text{int}}$ [$10^{-2}$ pN/nm] & $2.9\pm 0.3$ & $3.1\pm 0.4$ & $4.2\pm 0.4$ & $5.4\pm 0.3$ & $5.4\pm 0.3$  \\[1.5pt]
\hline\\[-2ex]
$\epsilon$ [$10^{-1}$ pN] & $3.0\pm 0.6$ & $2.5\pm 0.2$ & $3.4\pm 0.5$ & $4.4\pm 0.2$ & $4.0\pm 0.5$ \\[1.5pt]
\hline\\[-2ex]
$\tau$ [s] & $3\pm 1$ & $2.6\pm 0.7$ & $0.5\pm 0.5$ & $0.022\pm 0.003$ & $0.030\pm 0.007$ \\[1.5pt]
\hline\\[-2ex]
$\sigma$ [$\text{k}_{B}\text{T/s}$] & $0.7\pm 0.7$ & $3\pm 2$ & $8\pm 2$ & $18\pm 2$ & $19\pm 6$ \\[1.5pt]
\hline
\end{tabular}}
\vskip 5mm
    \centering
\tcr{\begin{tabular}{l|ccccc}
Force &  10 pN & 15 pN & 20 pN & 25 pN & 30 pN \\[1.5pt]
\hline\\[-2ex]
$k_x$ [$10^{-2}$ pN/nm] & $7.8\pm 0.4$ & $8.0\pm 0.5$ & $7.9\pm 0.4$ & $7.5\pm 0.5$ & $7.9\pm 0.5$\\[1.5pt]
\hline\\[-2ex]
$\mu_x$ [$10^4$ nm/(pN s)] & $2.3\pm 0.1$ & $2.4\pm 0.1$ & $2.6\pm 0.1$ & $2.6\pm 0.1$ & $2.8\pm 0.1$\\[1.5pt]
\hline\\[-2ex]
$k_y$ [$10^{-1}$ pN/nm]  & $1.1\pm 0.1$ & $1.3\pm 0.1$ & $1.1\pm 0.1$ & $1.0\pm 0.1$ & $0.9\pm 0.1$\\[1.5pt]
\hline\\[-2ex]
$\mu_y$ [$10^3$ nm/(pN s)] & $9\pm 1$ & $9\pm 2$ & $9\pm 1$ & $8\pm 1$ & $8\pm 1$\\[1.5pt]
\hline\\[-2ex]
$k_{\text{int}}$ [$10^{-2}$ pN/nm] & $5.0\pm 0.4$ & $5.3\pm 0.3$ & $5.0\pm 0.4$ & $4.6\pm 0.5$ & $3.8\pm 0.3$  \\[1.5pt]
\hline\\[-2ex]
$\epsilon$ [$10^{-1}$ pN] & $3.6\pm 0.3$ & $4.2\pm 0.3$ & $3.6\pm 0.3$ & $3.5\pm 0.2$ & $3.8\pm 0.2$ \\[1.5pt]
\hline\\[-2ex]
$\tau$ [$10^{-2}$ s] & $2.3\pm 0.2$ & $2.4\pm 3.7$ & $0.8\pm 0.5$ & $4\pm 1$ & $6\pm 2$ \\[1.5pt]
\hline\\[-2ex]
$\sigma$ [$\text{k}_{B}\text{T/s}$] & $18\pm 3$ & $26\pm 6$ & $15\pm 3$ & $13\pm 3$ & $12\pm 3$ \\[1.5pt]
\hline
\end{tabular}}

\vskip 5mm
    \centering

\tcr{\begin{tabular}{l|cccccc}
$k_b$ [$10^{-2}$pN/nm] & 0.12 & 0.35 & 0.52 & 0.9 & 1.3 \\[1.5pt]
\hline\\[-2ex]
$\mu_x$ [$10^4$ nm/(pN s)]   &  $4.2\pm 7$ &  $3\pm1$ &  $2.4\pm0.3$ &  $1.7\pm0.2$ &  $1.55\pm0.03$ \\[1.5pt]
\hline\\[-2ex]
$k_y$ [$10^{-2}$ pN/nm]   &  $9\pm2$ &  $12\pm0.7$ &  $13\pm 1$ &  $10\pm2$ &  $12.0\pm0.4$ \\[1.5pt]
\hline\\[-2ex]
$\mu_y$ [$10^3$ nm/(pN s)] &  $(3 \pm 1)\cdot10^2$ &  $3\pm1$ &  $4\pm 1$ &  $3.7\pm0.5$ &  $2.7\pm0.4$ \\[1.5pt]
\hline\\[-2ex]
$k_{\text{int}}$ [$10^{-2}$ pN/nm] &   $1.6\pm0.8$ &  $4\pm 2$ &  $3.6\pm 0.2$ &  $3.7\pm0.2$ &  $3.6\pm0.6$ \\[1.5pt]
\hline\\[-2ex]
$\epsilon$ [$10^{-1}$ pN] &  $4\pm1$ &  $2\pm1$ &  $0.60\pm0.03$ &  $0.4\pm0.1$ &  $0.38\pm0.01$ \\[1.5pt]
\hline\\[-2ex]
$\tau$ [$10^{-1}$ s]     &  $0.42\pm0.05$ &  $0.4\pm0.2$ &  $0.5\pm0.1$ &  $7\pm6$ &  $1.5\pm0.7$ \\[1.5pt]
\hline\\[-2ex]
$\sigma$ [$\text{k}_{B}\text{T/s}$]  &  $(3\pm1)\cdot 10^3$ &  $(1.0\pm0.7)\cdot 10^3$ &  $26\pm8$ &  $10\pm 8$ &  $4\pm2$ \\[1.5pt]
\hline\\[-2ex]
\end{tabular}}

\caption{\tcr{\textbf{Fit parameters for OT-stretching experiments.} (Top and middle) Fitting parameters for the results of stretched RBCs at different pulling forces at maximum trap stiffness, $k_b\sim 5\cdot 10^{-2}$. (Bottom) Fitting parameters for the results of Figure \ref{fig:FIG4}D at different trap stiffness. We show parameters averaged over 8 RBCs for each force (top and middle) and over 2-5 RBCs (bottom) with their statistical errors.}} 
\label{tab:2}
\end{table}

\clearpage

\begin{table}[bth!]
    \centering
    \small
\tcr{\begin{tabular}{l|cccccccc}
        RBC & P1 & P2 & H1 & H2 & H3 & H4 & H5\\[1.5pt]
        \hline\\[-2ex]
        $k_x$ [$10^{-3}$ pN/nm] & 6.0$\pm$0.7 & 8.1$\pm$0.5 & 6.5$\pm$0.6 & 15.1$\pm$0.2 & 3.7$\pm$0.6 & 7.17$\pm$0.03 & 9.5$\pm$0.4 \\[1.5pt]
        \hline\\[-2ex]
        $\mu_x$ [$10^4$ nm/(pN s)] & 2.6$\pm$0.3& 2.1$\pm$0.4& 2.8$\pm$0.5& 3.6$\pm$0.4 & 2.6$\pm$0.2 & 1.6$\pm$0.1 & 1.7$\pm$0.2 \\[1.5pt]
        \hline\\[-2ex]
        $k_y$ [$10^{-2}$ pN/nm] & 1.7$\pm$0.5 & 29$\pm$8 & 1.6$\pm$0.2 & 1.6$\pm$0.1 & 2.9$\pm$0.5 & 1.30$\pm$0.05 & 1.78$\pm$0.06\\[1.5pt]
        \hline\\[-2ex]
        $\mu_y$ [$10^4$ nm/(pN s)] & 140$\pm$10 & 31$\pm$8 & 2.9$\pm$0.4 & 2.3$\pm$0.3 & 0.57$\pm$0.08 & 1.96$\pm$0.02 & 1.94$\pm$0.02 \\[1.5pt]
        \hline\\[-2ex]
        $k_{\text{int}}$ [$10^{-3}$ pN/nm] & 1.8$\pm$0.6 & 6$\pm$2 & 4.5$\pm$0.5 & 4.2$\pm$0.2 & 2.3$\pm$0.4 &  2.88$\pm$0.05 & 2.84$\pm$0.06 \\[1.5pt]
        \hline\\[-2ex]
        $\epsilon$ [pN] & 0.7$\pm$0.1 & 2.1$\pm$0.5 & 2.6$\pm$0.1 & 2.4$\pm$0.1 & 4.7$\pm$0.3 & 3.86$\pm$0.02 & 4.41$\pm$0.03 \\[1.5pt]
        \hline\\[-2ex]
        $\tau$ [$10^{-2}$ s] & (2$\pm$1)$\cdot 10^2$ & (4$\pm$1)$\cdot 10^2$ & 8$\pm$1 & 1.1$\pm$0.1 & 2.5$\pm$0.3 & 17$\pm$3 & 7.7$\pm$0.1 \\[1.5pt]
        \hline\\[-2ex]
        $\sigma$ [$10^2$ $\text{k}_{B}\text{T/s}$] & 1.3$\pm$0.2 & 0.8$\pm$0.1 & 16$\pm$2 & 64$\pm$3 & 10$\pm$2 & 22$\pm$4 & 39$\pm$5 \\[1.5pt]
        \hline
\end{tabular}}
\normalsize
\caption{\textbf{Fit parameters for OT-sensing experiments}. Here, H stands for healthy RBC, whereas P stands for passivated RBCs. For each RBC we show the average estimate values with their statistical errors.}
    \label{tab:3}
\end{table}

\clearpage

\begin{table}[bth!]
    \centering
\footnotesize
\tcr{\begin{tabular}{l|ccccccc}
        RBC & P & H1 & H2 & H3 & H4 & H5 & H6\\[1.5pt]
        \hline\\[-2ex]
        $k_x$ [$10^{-2}$pN/nm] & 59$\pm$1 & 5.37$\pm$0.05 & 5.07$\pm$0.06 & 5.23$\pm$0.07& 4.61$\pm$0.05& 5.39$\pm$0.06& 8.3$\pm$0.1\\[1.5pt]
        \hline\\[-2ex]
        $\mu_x$ [$10^4$nm/(pN$\,$s)] & 0.8$\pm$0.1 & 1.2$\pm$0.1 & 1.3$\pm$0.2 & 1.2$\pm$0.2 & 1.4$\pm$0.2 & 1.2$\pm$0.2 & 0.9$\pm$0.2\\[1.5pt]
        \hline\\[-2ex]
        $k_y$ [$10^{-2}$pN/nm] & 5.6$\pm$0.2 & 1.93$\pm$0.03 & 1.89$\pm$0.03 & 1.92$\pm$0.05 & 1.95$\pm$0.03 & 2.10$\pm$0.03 & 1.35$\pm$0.02\\[1.5pt]
        \hline\\[-2ex]
        $\mu_y$ [$10^4$nm/(pN$\,$s)] & 0.082$\pm$0.003 & 4.85$\pm$0.02 & 4.89$\pm$0.03 & 4.79$\pm$0.06 & 4.84$\pm$0.03 & 4.95$\pm$0.04 & 4.65$\pm$0.02\\[1.5pt]
        \hline\\[-2ex]
        $k_{\text{int}}$ [$10^{-3}$pN/nm] & 91$\pm$3 & 3.8$\pm$0.1 & 3.57$\pm$0.05 & 3.8$\pm$0.1 & 3.50$\pm$0.05 & 3.94$\pm$0.08 & 4.16$\pm$0.05\\[1.5pt]
        \hline\\[-2ex]
        $\epsilon$ [pN] & 0.2$\pm$0.1 & 4.42$\pm$0.02 & 4.37$\pm$0.04 & 4.43$\pm$0.02 & 4.40$\pm$0.06 & 4.45$\pm$0.01 & 4.52$\pm$0.05\\[1.5pt]
        \hline\\[-2ex]
        $\tau$ [$10^{-2}$s] & 750$\pm$10 & 6$\pm$2 & 4.1$\pm$0.5 & 6$\pm$1 & 3.9$\pm$0.01 & 3.3$\pm$0.3 & 7$\pm$1\\[1.5pt]
        \hline\\[-2ex]
        $\sigma$ [$10^3\,\text{k}_{B}\text{T/s}$] & (5.7$\pm$0.3)$\cdot 10^5$ & 7.13$\pm$0.09 & 6.96$\pm$0.09 & 7.2$\pm$0.2 & 6.5$\pm$0.2 & 6.6$\pm$0.1 & 7.4$\pm$0.4 \\[1.5pt]
        \hline
        \\[-2ex]
        $\mathcal{V}_{x}$ [$10$ nm] & 1.4$\pm$0.3 & 32$\pm$2 & 40$\pm$2& 42$\pm$2 & 47$\pm$2 & 35$\pm$2 & 42$\pm$3\\[1.5pt]
        \hline
\end{tabular}}
\normalsize
\caption{\textbf{Fit parameters for OM experiments.}  H stands for healthy RBCs, whereas P stands for passivated RBC. For a given RBC, we show parameter values averaged over all 512 points along the RBC contour (Fig.~\ref{fig:RBC1-5}) together with their statistical error.}
    \label{tab:4}
\end{table}

\begin{table}[bth!]
    \centering
    \small
    \tcr{
\begin{tabular}{l|cc}
        Simulation & P & A \\[1.5pt]
        \hline\\[-2ex]
        $k_x$ [$10^{-5}$ pN/nm] & 8.6 $\pm$ 0.1 & 2.31 $\pm$ 0.01  \\[1.5pt]
        \hline\\[-2ex]
        $\mu_x$ [$10^7$ nm/(pN s)] & 5.66 $\pm$ 0.06 & 4.86 $\pm$ 0.02  \\[1.5pt]
        \hline\\[-2ex]
        $k_y$ [$10^{-1}$ pN/nm] & 2 $\pm$ 1 & 2.9 $\pm$ 0.6  \\[1.5pt]
        \hline\\[-2ex]
        $\mu_y$ [$10^3$ nm/(pN s)]  & 982 $\pm$ 3 & 9.01 $\pm$ 0.03  \\[1.5pt]
        \hline\\[-2ex]
        $k_{\text{int}}$ [$10^{-4}$ pN/nm] & 3 $\pm$ 1 & 2.41 $\pm$ 0.06 \\[1.5pt]
        \hline\\[-2ex]
        $\epsilon$ [pN] & 0 $\pm$ 1 &3 $\pm$ 1 \\[1.5pt]
        \hline\\[-2ex]
        $\tau$ [$10^{-2}$ s] & 6 $\pm$ 5 & 0.53 $\pm$ 0.03  \\[1.5pt]
        \hline\\[-2ex]
        $\sigma$ [$\text{k}_{B}\text{T/s}$] & 29 $\pm$ 60 & (1.0 $\pm$ 2)$\cdot$ $10^4$    \\[1.5pt]
        \hline
\end{tabular}}

\normalsize
\caption{\textbf{Fitting parameters for simulations of OT-sensing experiments}. Here, P stands for passive simulations, whereas A stands for active simulations. The fitting algorithm estimates uncertainties. } 
    \label{tab:5}
\end{table}

\end{document}